\documentclass[prd,reprint]{revtex4-1}
\usepackage{blindtext}
\usepackage{amsfonts}
\usepackage{physics}
\usepackage{graphicx}
\usepackage{caption}
\usepackage{subcaption}
\graphicspath{{./Plots/}}
\usepackage[export]{adjustbox}
\usepackage{float}
\usepackage{xcolor} %DA RIMUOVERE

\usepackage{multirow}

\begin{document}

\title{The fate of high winding number topological phases in the disordered extended Su-Schrieffer-Heeger model}
 
\author{Emmanuele G. Cinnirella$^{(1,2)}$, Andrea Nava$^{(3)}$, Gabriele Campagnano$^{(4)}$, and Domenico Giuliano$^{(1,2)}$}
\affiliation{
$^{(1)}$Dipartimento di Fisica, Universit\`a della Calabria Arcavacata di 
Rende I-87036, Cosenza, Italy \\
$^{(2)}$I.N.F.N., Gruppo collegato di Cosenza, 
Arcavacata di Rende I-87036, Cosenza, Italy \\
$^{(3)}$Institut f\"ur Theoretische Physik IV, Heinrich-Heine-Universit\"at, 40225 D\"usseldorf, Germany \\
$^{(4)}$CNR-SPIN,  c/o Complesso di Monte S. Angelo, via Cinthia - 80126 - Napoli, Italy}

\begin{abstract}
We use the Lindblad equation approach to investigate topological phases hosting more than one localized state at each side of a disordered SSH chain with properly tuned long range hoppings. Inducing a non equilibrium steady state across the chain, we probe the robustness
of each phase and the fate of the edge modes looking at the distribution of electrons along the chain and the corresponding standard deviation in the presence of different kinds of disorder either preserving, or not, the symmetries of the Hamiltonian.
\end{abstract}
\date{\today}
\maketitle
%\author{Me}
%\email{mail@example.com}
%\author{Myself}
%\author{Someone Else}
%\affiliation{A University}

%\begin{abstract}
%Here I tell what I have done... And I have done a lot but it is hard to tell what exactly I have done...
%\end{abstract}

\maketitle

\section{Introduction}
\label{intro}
Edge states in one dimensional (1D) systems promise to play a crucial role in quantum computation. Due to their unique properties that can be exploited for qubit manipulation, error correction, and braiding operations, the edge states hold the promise of increased stability and fault tolerance, which are critical challenges in the development of quantum algorithms.
Proper materials and setups, which can host and manipulate the edge states, are thus required in topological quantum computations. Such states can be realized in a wide class of systems, from  anyons with non-Abelian statistics realized for example in quantum Hall states at filling fraction $5/2$ \cite{Moore1991}, helical electron liquids \cite{Oreg2010}, or semiconductor-superconductor heterostructures \cite{Lutchyn2010,Alicea2010,Sarma2015}, to helical optical states in photonic metamaterials \cite{Mousavi2015,Parappurath2020} or even in topological mechanical systems \cite{Chen2016,Upadhyaya2020}.

\begin{figure}
\includegraphics[width=\linewidth]{{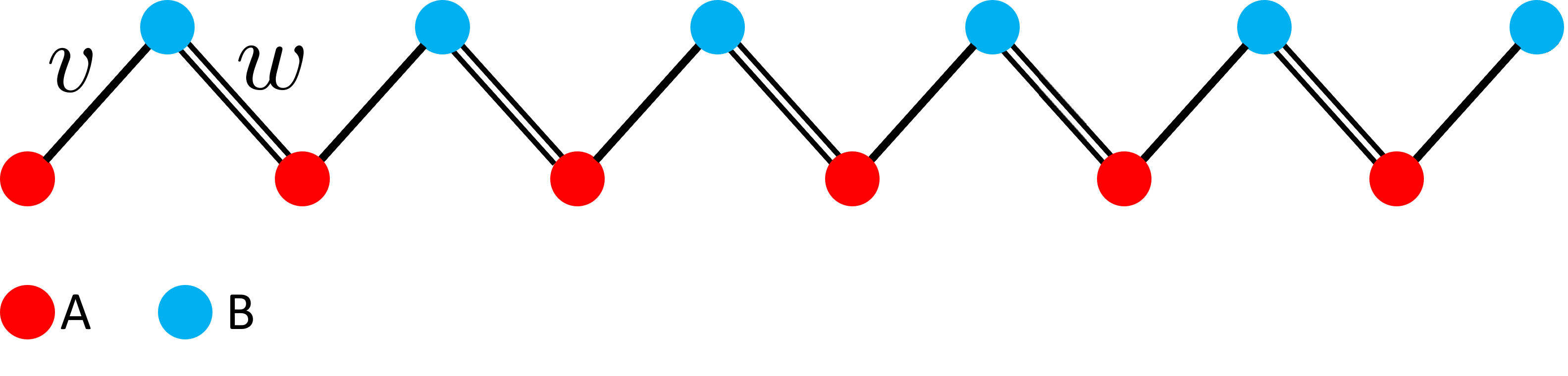}}
\caption{Pictorial representation of the SSH chain. Single and double solid lines represent the intra- and inter-dimer hopping, respectively.}
\label{fig:SSH_normal}
\end{figure}

Among all, the simplest non trivial 1D system that manifests topological edge states and disorder tolerance is the Su–Schrieffer–Heeger (SSH) model. It consists of a noninteracting tight-binding model of connected dimers. Such a model, introduced for the first time in 1979 to describe the transport properties of polyacetylene \cite{Su1979}, is the nonsuperconducting analog of the Kitev chain \cite{Kitaev2001}, thus experimentally and theoretically more accessible, hosting Dirac rather than Majorana modes at the boundary.
In the SSH model, the topological transition is controlled by tuning the ratio of the hoppings between two consecutive odd-even (intra-dimer) sites and even-odd (inter-dimer) ones (single and double lines in Fig.\ref{fig:SSH_normal} respectively). When the intra-dimer hopping strength is weaker (stronger) than the inter-dimer one, the system exhibits two (zero) edge modes exponentially localized at the boundary of the system with open boundary conditions. This property is shared by all the SSH Hamiltonians adiabatically connected to each other.
On the other side, in the presence of periodic boundary conditions, a topological invariant, like the Chern number, defined as the integral of the Berry curvature over the Brillouin zone of the system \cite{Berry1984}, can be introduced to discriminate between different topological phases. This quantity, which is quantized and assumes the same value for adiabatically connected Hamiltonians, in the SSH model can only take the value zero or one. As long as the chiral, time reversal and particle-hole symmetries are preserved, the SSH model belongs to the BDI class of the Altland-Zirnbauer classification of topological insulators \cite{Altland1997}, and this ensures that the 'bulk-boundary correspondence' is valid \cite{Fidkowski2011,Prodan2016,Chen2020}, i.e., the topological invariant defined in the translational invariant system corresponds to the number of edge states at each boundary of the corresponding system with open boundary conditions. The SSH model can be experimentally realized in cold atoms systems like bosonic lattice gas \cite{Grusdt2013}, Rydberg synthentic lattice of $^84$Sr \cite{Kanungo2022} and $^87$Rb \cite{Meier2018} or also in optical waveguide \cite{Wang2022,Kang2023} and photonic quantum walk setups \cite{Cardano2016,Cardano2017,Derrico2020} in a way that can be manipulated to perform quantum information encoding in dot arrays \cite{Petropoulos2022} and quantum braiding in Y-junction gates \cite{Boross2019}.

Recently, great interest has been attracted by generalized versions of the SSH model, called extended SSH models (eSSH) \cite{Li2019,Gonzalez2019,Malakar2023}. In this class of systems, the hopping between even or odd sites, as well as the presence of a nonzero onsite energy, breaks particle-hole and chiral symmetry. The bulk-boundary correspondence is no longer valid, i.e., the one-to-one correspondence between the topological invariant and the number of edge states in the open system is lost \cite{Li2019}. On the contrary, the chiral symmetry - and the bulk-boundary correspondence -  are preserved for long range hopping connecting odd sites with even ones. In this case, the winding number is no longer forced to assume only two values, as in the standard SSH model, but new topological phases with more than one edge state at each boundary (that increase with the range of the hopping) are supported. On the experimental side, such systems could be experimentally realized by applying pertinently fine tuned high-frequency ac-driving fields on an SSH chain \cite{Gonzalez2019}, in optically resonant nanoparticles \cite{Li2019} or in photonic crystal systems \cite{Li2014}. The possibility to tune more than one edge states could be a crucial ingredient in quantum computation.

Clearly, real systems are less than perfect, and it is extremely important to be able to predict how much these new topological phases, characterized by more than one edge states, are robust against different types of noise and defects. The simultaneous presence of topology and disorder has always attracted a lot of interest due to non trivial effects that can emerge when both these ingredients are present \cite{Meier2018}. While topological phases are generally robust to certain types of disorder up to some characteristic strengths, topological features can be totally faded away or even enhanced, inducing a reentrant topological phase transition at larger values of the disorder strength \cite{Pientka2013,Nava2017,Zuo2022}. Furthermore, while uncorrelated disorder is expected to induce Anderson localization \cite{Anderson1958,Abrahams1979,Evers2008}, correlated disorder can allow for the existence of delocalized states which in turn influences the behavior of the boundary states \cite{Aoki1983,Phillips1991,Phillips1991b,Wu1992,Sedrakyan2004}. It is worth noting that non trivial types of correlated disorder are experimentally accessible by means, for example, of photoluminescence and vertical dc resistance \cite{Bellani1999} or in ultracold atoms \cite{Schaff2010,Billy2008,Sanchez-Palencia2010} and photonic systems \cite{Dietz2011}. 

In this paper, we investigate the robustness of eSSH models, hosting more than one edge state, in the presence of different types of disorder than can break or preserve the symmetries of the clean system. We make use of the Lindblad master equation (LE) formalism \cite{Lindblad1976} describing the Markovian dynamics of the density matrix of the system when it is coupled to the environment (i.e., the bath). 
In recent years, the LE has been successfully applied to different contexts, from ultracold atoms \cite{Braaten2017,Lee2019} to 
 condensed matter systems \cite{Lieu2020, Budich2015,Nava2019,Nava2022_2,Nava2022_3}, quantum biology and quantum chemistry \cite{Plenio2008, Manzano2013, Pino2015, Nava2022} or to implement algorithms for quantum and classical problems \cite{Nurdin2009,Verstraete2009,Diehl2008,Schlimgen2022,Nava2022_4,Nava2023,Bagarello2022,Bagarello2023}.
Furthermore, the Lindblad approach has recently  been used to study (dynamical) topological phase transitions in one- and two-dimensional systems \cite{Diehl2011,Goldstein2019,Shavit2020,Beck2021} and planar superconductors \cite{Cui2019,nava2023L,nava2023S} as well as the manybody localization in interacting systems \cite{thompson2023,artiaco2023,artiaco2023_2}. It has also been implemented to investigate both relaxation dynamics toward a thermal state
 \cite{Tarantelli2022,DiMeglio2020, Leeuw2021,Dangel2018,Artiaco2021,deleeuw2023hidden}, as well as the non-equilibrium steady states (NESSs) that emerge when a system is placed in contact with two reservoirs at different temperatures or voltage bias/chemical potentials \cite{Guimaraes2016,Tarantelli2021,Popkov2017,Iztok2013,Benenti2009b,Benenti2009,Nava2021,Maksimov2022}.

In the LE formalism, after tracing out the bath degrees of freedom, the interaction between the system and the bath is modeled in terms of “jump” operators. Here, we consider an eSSH model connected to two reservoirs at its endpoints to drive the system towards an out-of-equilibrium configuration injecting, or removing, particles through its boundaries. When working in the large bias limit, one of the reservoirs acts as an electron "source" while the other plays the role of an electron "drain" \cite{Benenti2009,Benenti2009b,Nava2021,Nava2022}. After a transient regime, the system reaches the NESS characterized by time independent current along the chain and site dependent real space density through which the topological properties of the system can be investigate. Indeed, we implement the even-odd occupancy (EOD), i.e., the difference between the mean occupation on the even and odd sites \cite{nava2023_ssh}, as a topological invariant. The EOD allows us to monitor the nontrivial topological properties of the disordered eSSH and to map out the full disorder dependent phase diagram. This procedure circumvents the limitations of alternative numerical and analytical approaches, like the disordered averaged winding number (DAWN) \cite{Mondragon2014,Lin2021} or the strong disorder renormalization group (SDRG) \cite{Fisher1992,Fisher1994,Fisher1995,Ma1979,Dasgupta1980}, and can be experimentally measured in out-of-equilibrium experiments \cite{nava2023_ssh}. 

Using the EOD we investigate the phase diagram of the eSSH model as a function of disorder, also performing a comparison with analytical results obtained within the SDRG approach within appropriate limits. We show that a sort of hierarchy is observed in the way disorder destroys topological phases characterized by a high value of the topological invariant. Increasing the disorder strength, the topological invariant is reduced through unitary steps, via the appearance of disorder induced "buffer" phases, rather than an abrupt transition toward the topologically trivial phase hosting no zero energy modes. At the same time, disorder can
lead to reentrant topological phases in favor of phases hosting a single zero energy mode at each boundary, like that as observed in the standard SSH model or in the Kitaev chain \cite{Pientka2013,Nava2017,nava2023_ssh}. Monitoring the standard deviation of the EOD, after computing its average over many disorder configurations, as a function of disorder strength and length of the chain, we can identify the Griffiths effect that takes place in a narrow area around each phase transition \cite{Motrunich2001,Sau2013} and also to distinguish it from other mimicking effects that take place in the presence of disorder that breaks the chiral symmetry of the system. In summary, we argue that a simultaneous comparison of the EOD and its standard deviation allows us to characterize the properties of the eSSH model in the presence of disorder in order to predict the robustness of the zero energy modes. 
%and to determine for which range of the parameter space a given phase is suitable to be implemented in a quantum computation setup.

The paper is organized as follows:
\begin{itemize}
    \item In Section \ref{sec:model} we introduce the model Hamiltonian for different families of eSSH chains and review the LE approach as well as the definition of the EOD.
    \item In Section \ref{sec:noise} we introduce the different types of disorder analyzed in the paper and the adopted numerical procedure.
    \item In Section \ref{sec:SDRG} we implement the SDRG approach to gain some insight into on the boundaries of each topological phase in the presence of disorder.
    \item In Section \ref{sec:numerical_results} we discuss the main numerical results for different types of disorder and eSSH models. We also investigate the EOD, its standard deviation and the area associated with each topological phase as a function of disorder strength.
    \item In Section \ref{sec:conclusions} we summarize and comment on our results and provide possible further developments of our work.
    \item In the Appendix, we review the SDRG approach and derive the recursive equations for a generic long range eSSH model.
\end{itemize}

\section{Model and methods}
\label{sec:model}
The standard SSH chain is a 1D lattice model constituted by a periodic repetition of $N$ two-site unit cells, the dimer. The $L=2N$ spinless sites  can be bipartited into two sublattice consisting of the first ($A$) and second ($B$) sites of each dimer respectively, as shown in Fig.\ref{fig:SSH_normal}. 

The SSH model is defined through the Hamiltonian
\begin{equation}
    H_{v,w}=\sum_{j=1}^{N}\left( vc^\dagger_{A,j}c_{B,j}+w c^\dagger_{B,j} c_{A,j+1}\right)+\text{h.c.}
    \label{eq:ssh_normal}
\end{equation}
In Eq.\eqref{eq:ssh_normal} we denote by $c^\dagger_{X,i}$ and $c_{X,i}$ the creation and annihilation operators for a spinless electron on dimer $i$ and sublattice $X=A,B$, satisfying the standard anticommutation relations
\begin{align}
    \{c^\dagger_{X,i},c_{Y,j}\} & =\delta_{XY}\delta_{ij} \nonumber \\
    \{c^\dagger_{X,i},c^\dagger_{Y,j}\} & =\{c_{X,i},c_{Y,j}\}=0
\end{align}
With $v$ and $w$ we denote the intra- and inter- dimers hopping strength, respectively. The SSH chain is the simplest 1D model presenting topological behavior as a function of the ratio $v/w$. 

The system exhibits a gapped spectrum except at $v=w$, where the topological transition takes place. The two phases can be distinguished in the presence of open boundary condition where, for $v<w$, the energy spectrum of the Hamiltonian displays two zero-energy modes, associated with two eigenstates that are exponentially localized at the first and last sites of the chain while, for $v>w$, the gap is totally empty. The SSH model belongs to the BDI class of the Altland-Zirnbauer classification of topological insulators, characterized by having particle-hole, time-reversal, and chiral (or sublattice) symmetry. Defining the chiral operator as
\begin{equation}
    \Gamma=\sum_j\left( c^\dagger_{A,j}c_{A,j}-c^\dagger_{B,j}c_{B,j}\right)
\end{equation}
the Hamiltonian satisfies the relation $\{\Gamma,H\}=0$ that implies a symmetric spectrum around the zero, i.e., each eigenstate has a chiral partner at opposite energy. Due to this symmetry, a bulk-edge correspondence can be established for the SSH model, i.e., an integer values topological  invariant that can be defined in the presence of periodic boundary conditions, which corresponds to the number of edge states located at the boundaries of the open chain.
Indeed, by imposing periodic boundary conditions on Eq.\eqref{eq:ssh_normal}, we can relate the number of edge states to the Chern number defined by means of the bulk eigenstates. In particular, in 1D the Chern number corresponds to the Zak phase, i.e., to the integral of the Berry connection over a closed path throughout the whole Brillouin zone.
%\begin{equation}
%    \omega=\frac{1}{i\pi}\oint_{BZ} dk \bra{\psi_{k,-}}\nabla_k\ket{\psi_{k,-}}=
%    \begin{cases}
%    1 &\text{ if } w>v\\
%    0 &\text{ if } w<v \\
%    \text{undef.} &\text{ if } v=w 
%    \end{cases}
%    \label{eq:winding}
%\end{equation}
%With $\ket{\psi_{k,-}}$ indicating the eigenstates having wavevector $k$ and belonging to the lowest energy band.
\noindent
Writing the Hamiltonian in momentum space using
\begin{equation}
c_{X,j}=\frac{1}{\sqrt{N}}\sum_{k}e^{ikj}c_{X,k}
\end{equation}
for $X=A,B$, we can set
\begin{equation}
H_{v,w}=\sum_{k}(\begin{array}{cc}
c_{A,k}^{\dagger} & c_{B,k}^{\dagger}\end{array})H(k)\left(\begin{array}{c}
c_{A,k}\\
c_{B,k}
\end{array}\right)
\end{equation}
where
\begin{equation}
%    H(k)=h_x(k)\sigma_x+h_y(k)\sigma_y
     H(k)=\gamma(k) \cdot \overrightarrow{\sigma}
     \label{eq:gamma}
\end{equation}
with $\sigma_i$, $i=x,y,z$, being the Pauli matrices and ${\gamma(k)=(v+w \cos k,w \sin k,0)}$. The Zak phase corresponds to the winding number, $\omega$, of the closed curve $\gamma(k)$ i.e., the number of times the closed curve revolves around the origin in the $\gamma_x-\gamma_y$ plane.
The bulk-edge correspondence remains valid also in the presence of long range hoppings and disorder that preserve the chiral symmetry. This is realized, for example, in the presence of hoppings that connect sites of the sublattices $A$ and $B$ at any distance but not in the presence of hopping between sites belonging to the same sublattice, with the Hamiltonian changing from the BDI class to the trivial AI class.

\subsection{Extended SSH models}\label{sec:eSSH_models}
\begin{figure}
\includegraphics[width=\linewidth]{{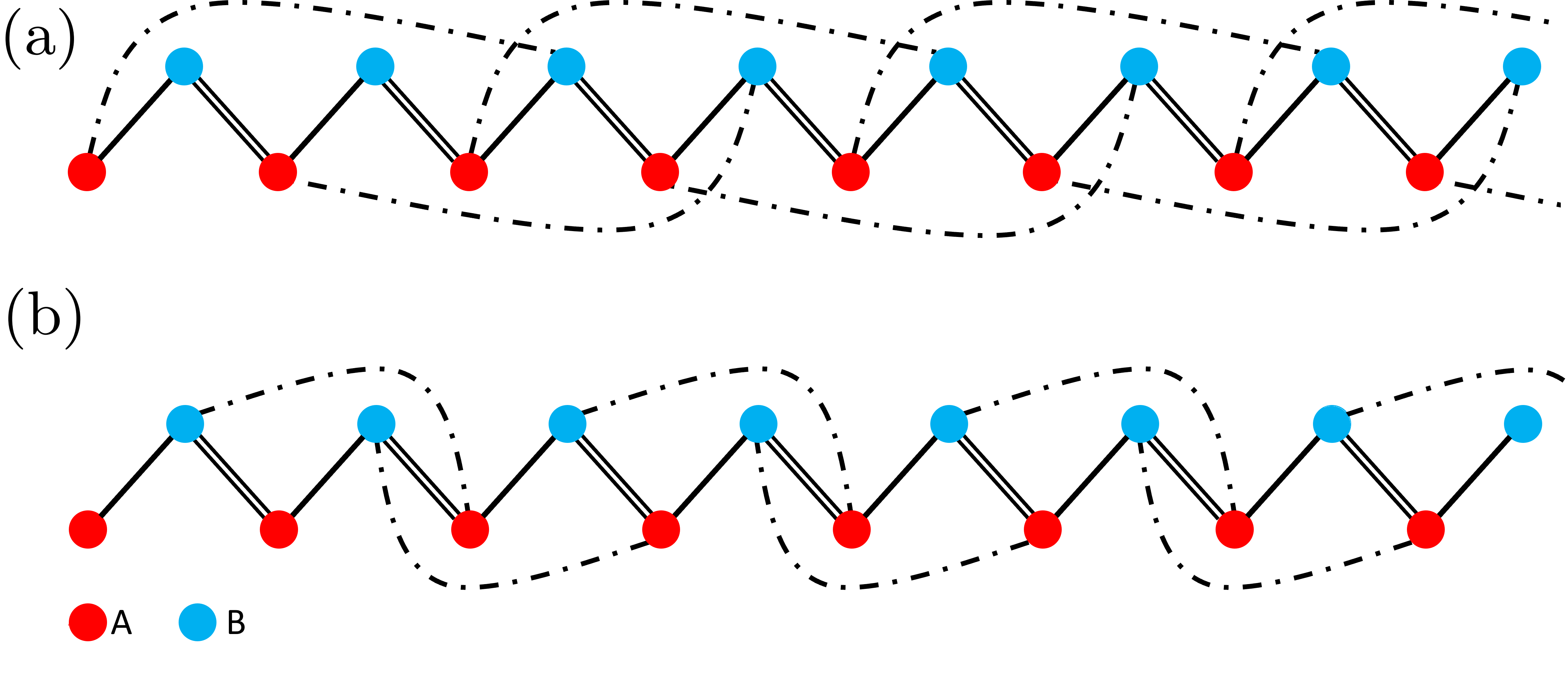}}
\caption{Pictorial representation of the eSSH for: a) $H^{\text{A-B}}_{2}$ and b) $H^{\text{B-A}}_{2}$. The long range hopping (dashed line) connects next-nearest-neighbor dimers. Red (blue) dots belong to the A (B) sublattice.}
\label{fig:SSH_extended}
\end{figure}

Introducing a long range hopping between the two sublattices of the SSH chain, it is possible to define a family of Hamiltonians, called extended SSH (eSSH) models \cite{Li2019,Gonzalez2019,Malakar2023}. These Hamiltonians exhibit high values of the winding number and, as a consequence, they can host more than one edge state at each boundary. Two families of chiral symmetric long range hopping Hamiltonians can be defined as
\begin{align}
H^{\text{A-B}}_{n}=H_{v,w} + \sum_j zc^\dagger_{A,j}c_{B,j+n}+\text{h.c.}
\label{eq:familyAB}
\\
H^{\text{B-A}}_{n}=H_{v,w} + \sum_j zc^\dagger_{B,j}c_{A,j+n}+\text{h.c.}
\label{eq:familyBA}
\end{align}
with $n$ the range of the hopping and $z$ the long range hopping strength (see Fig.\ref{fig:SSH_extended} for a pictorial representation of the chain for $n=2$).

\noindent

\noindent
Hamiltonian $H^{\text{A-B}}_{n}$ ($H^{\text{B-A}}_{n}$) gives rise to topological non trivial phases having winding number up to $n$ ($-n$), meaning that exactly $2|n|$ localized edge states are present in the set of one-particle eigenstates having zero energy, because of the bulk-edge correspondence.

The sign and the magnitude of $n$ dictate the number of edge states and on which sites they are localized. More explicitly, the value of $|n|$ determines the number of edge states localized at both boundaries of the eSSH chain. If $n$ is positive (negative), these edge states will be localized on the first $|n|$ sites of sublattice $A$ ($B$) and on the last $|n|$ sites of sublattice $B$ ($A$). 
Writing Eq.\eqref{eq:familyAB} and Eq.\eqref{eq:familyBA} in momentum space, we can introduce the closed loop $\gamma(k)$ as done in Eq.\eqref{eq:gamma} for the SSH model.
By looking at the behavior of the closed curve $\gamma(k)$ as a function of $v$, $w$, and $z$ it is easy lo locate the parameters boundaries corresponding to each topological phase, i.e., to each value of the winding number. As it happens for the simple SSH model, the parameter space of the eSSH chain splits into distinct regions, characterized by the same value of $\omega$ such that two Hamiltonians in the same phase are adiabatically connected to each other. All the Hamiltonians belonging to a given phase share the same topological properties as an appropriate limiting case, in which all the hoppings, except one, are sent to zero. Looking at these extreme cases it is clear why and where multiple edge states are expected in eSSH models. Let us consider the cases $H^{\text{A-B}}_{n}$ and $H^{\text{B-A}}_{n}$ and let us tune each hopping to zero, one-by-one.

Trivially, by sending $z\rightarrow 0$, we retrieve the standard SSH model, i.e., $H_{v,w}$, for both $H^{\text{A-B}}_2$ and $H^{\text{B-A}}_2$. Sending also $v \rightarrow 0$ gives rise to two zero-energy
Dirac fermions decoupled from the bulk and localized at the first $A$ site and at the last $B$ site. Viceversa, by sending $w \rightarrow 0$, the chain reduces to a collection of decoupled dimers with energies $\pm w$. It follows that
\begin{equation}
    z=0\Rightarrow\omega =
    \begin{cases}
    1 &\text{ if } v<w\\
    0 &\text{ if } v>w
    \end{cases}
\end{equation}

\noindent
\begin{figure}
\includegraphics[width=\linewidth]{{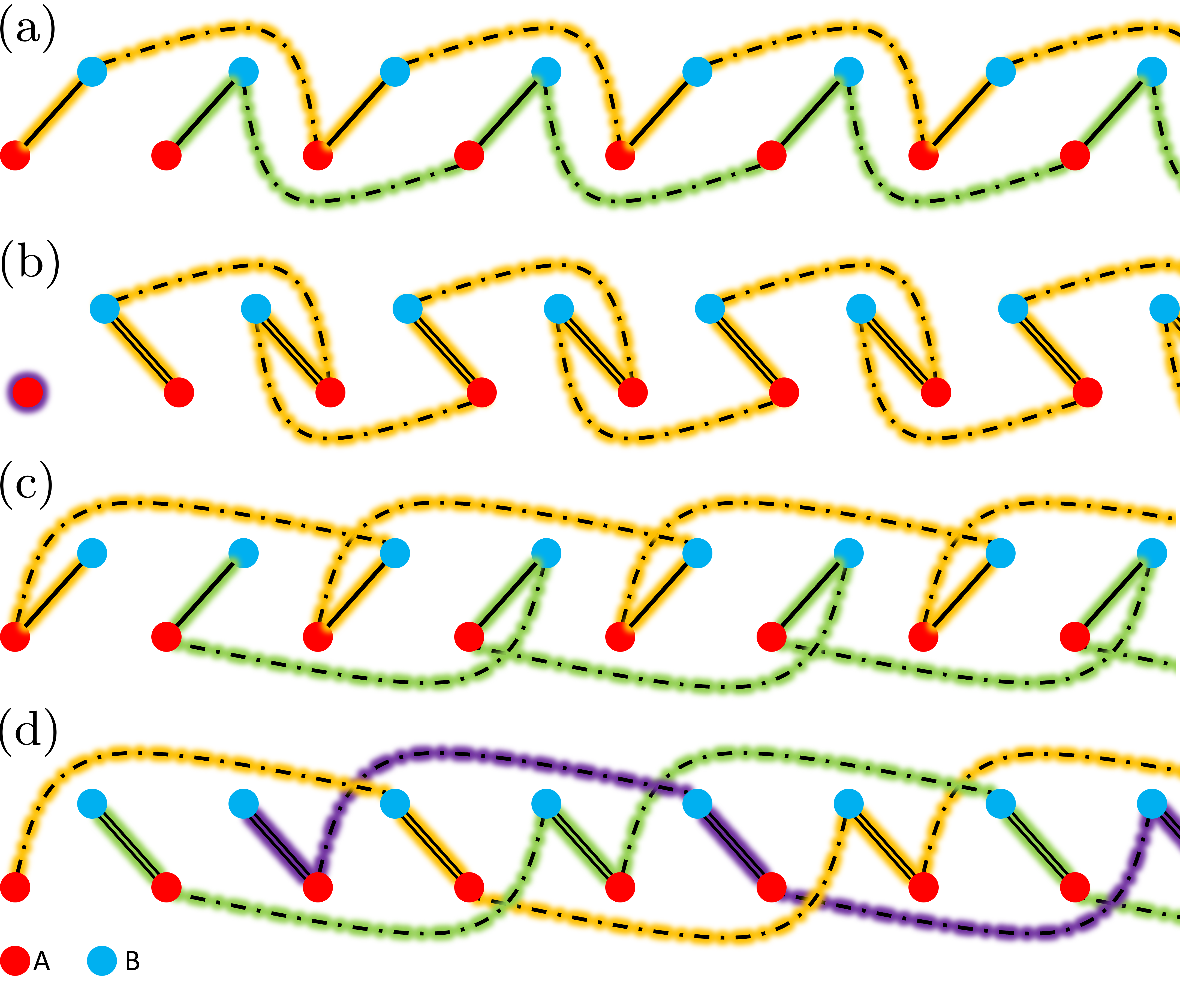}}
\caption{Pictorial representation of eSSH with one hopping sent to zero: a) $H^{\text{B-A}}_2$ for $w \rightarrow0$, b) $H^{\text{B-A}}_2$ for $v \rightarrow0$, c) $H^{\text{A-B}}_2$ for $w \rightarrow0$, d) $H^{\text{A-B}}_2$ for $v \rightarrow0$. In each panel, we have highlighted with different colors (yellow, green, and purple) each of the chains in which the eSSH model decouples.}
\label{fig:SSH_extreme}
\end{figure}

\noindent
For the cases in which $v$ or $w$ is the first hopping sent to zero, we have to analyze $H^{\text{A-B}}_2$ and $H^{\text{B-A}}_2$ separately.

\noindent
Let us start by sending first $w \rightarrow 0$ in $H^{\text{B-A}}_2$. After doing so, as shown in panel a) of Fig.\ref{fig:SSH_extreme}, even and odd dimers decouple in two disconnected SSH chains with intra-dimer hopping $v$ and inter-dimer hopping $z$. Clearly, when $v<z$ both chains are in the topological phase characterized by $\omega=1$ (with the edge states lying on the first, third, last and third to last sites of the chain) while for $v>z$ we have $\omega=0$. It follows that
\begin{equation}
    w=0\Rightarrow\omega =
    \begin{cases}
    0 &\text{ if } v>z\\
    2 &\text{ if } v<z
    \end{cases}
\end{equation}

\noindent
    On the other hand, sending $v\rightarrow0$ first, decouples the first and last Dirac fermions from the full chain, with the remaining sites rearranged in a SSH model with intra-dimer hopping $w$ and inter-dimer hopping $z$, as we show in panel b) of Fig.\ref{fig:SSH_extreme}. As a consequence, the total winding number is at least $\omega_{min}=1$ and increases further if $w<z$. Again, the zero energy states occupy the first, third, last and third to last sites. We have
\begin{equation}
    v=0\Rightarrow\omega =
    \begin{cases}
    1 &\text{ if } w>z\\
    2 &\text{ if } w<z
    \end{cases}
\end{equation}

\begin{figure}
\includegraphics[width=1 \linewidth]{{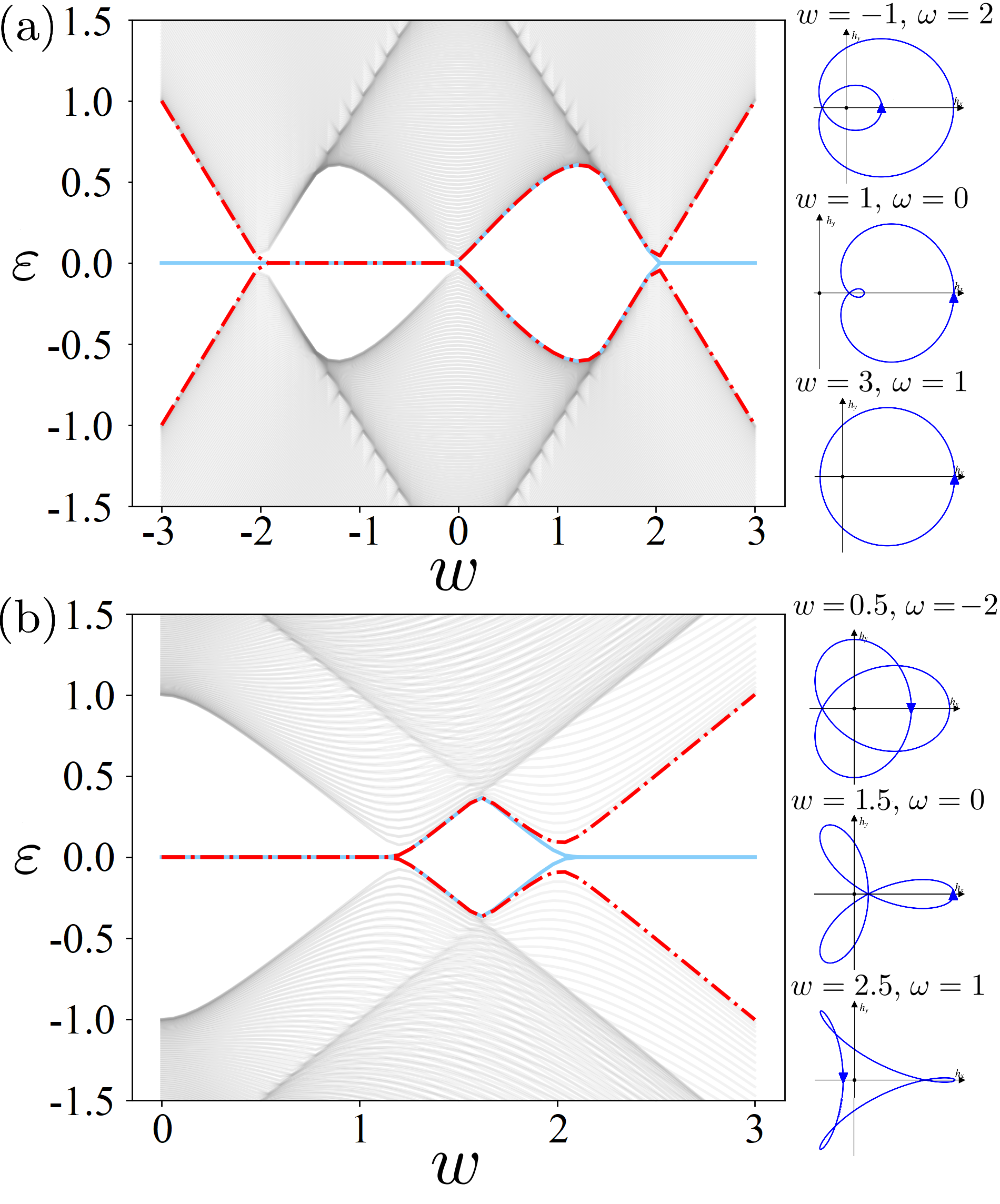}}
\caption{An example of the energy spectrum of: a) $H^{\text{B-A}}_2$ as a function of $w$ for $v=1+w/3$ and $z=1-w/3$, b) $H^{\text{A-B}}_2$ as a function of $w$ for $v=0.5$ and $z=1.5$. Continue blue and dashed red lines represent the four states closest to the center of the gap. The side panels show the behavior of $\gamma(k)$ for fixed values of the hoppings in each topological phase.}
\label{fig:energy_levels}
\end{figure}

\noindent
Let us move to $H^{\text{A-B}}_2$ and send $w \rightarrow 0$ first. As shown in panel c) of Fig.\ref{fig:SSH_extreme}, the system decouples in two SSH chains, but with reversed $A$ and $B$ sublattices. Again, the system can host two edge states on each boundary, this time located on the second, fourth, penultimate, and fourth to last sites of the full chain, when $v$ is lower than $z$. We can write
\begin{equation}
    w=0\Rightarrow\omega =
    \begin{cases}
    0 &\text{ if } v>z\\
    -2 &\text{ if } v<z
    \end{cases}
\end{equation}
\noindent
Finally, by sending $v\rightarrow0$ we decouple the eSSH model in a collection of three SSH chains. Looking at panel d) of Fig.\ref{fig:SSH_extreme}, the first chain has intra-hopping $z$ and inter-hopping $w$, while the other two chains have switched both the hopping and the sublattice index. It follows that, if $z<w$ ($z>w$), the first chain is in the nontrivial (trivial) topological phase with the other two chains in the trivial (nontrivial) one, so that
\begin{equation}
    v=0\Rightarrow\omega =
    \begin{cases}
    1 &\text{ if } w>z\\
    -2 &\text{ if } w<z
    \end{cases}
\end{equation}
Topology ensures that these properties are preserved even away from the extreme limits discussed above  if the initial and final Hamiltonians are adiabatically connected, i.e., the spectrum remains gapped. In Fig.\ref{fig:energy_levels} we show the energy spectrum of both models discussed in this Section, moving across a line in the full parameter space. In particular, in panel a) we show the eigenvalues of $H^{\text{B-A}}_2$ as a function of the inter-hopping strength for $v=1+w/3$ and $z=1-w/3$. The spectrum is always gapped except for $v=\{-2,0,2\}$ where, by increasing $w$, the topological phase transitions take place, moving from a phase with $|\omega|=1$ to a phase with $|\omega|=2$, then $\omega=0$ and again to $|\omega|=1$. In panel b) a similar behavior is shown for $H^{\text{A-B}}_2$ for $v=0.5$ and $z=1.5$. The sign of the winding number cannot be inferred from the eigenvalues alone; therefore, for this reason, on the right hand side of both panels, we show the closed curve $\gamma(k)$ in the $\gamma_x-\gamma_y$ plane.

\noindent

\subsection{Out-of-equilibrium even-odd differential occupation} 

\begin{figure}
\includegraphics[width=\linewidth]{{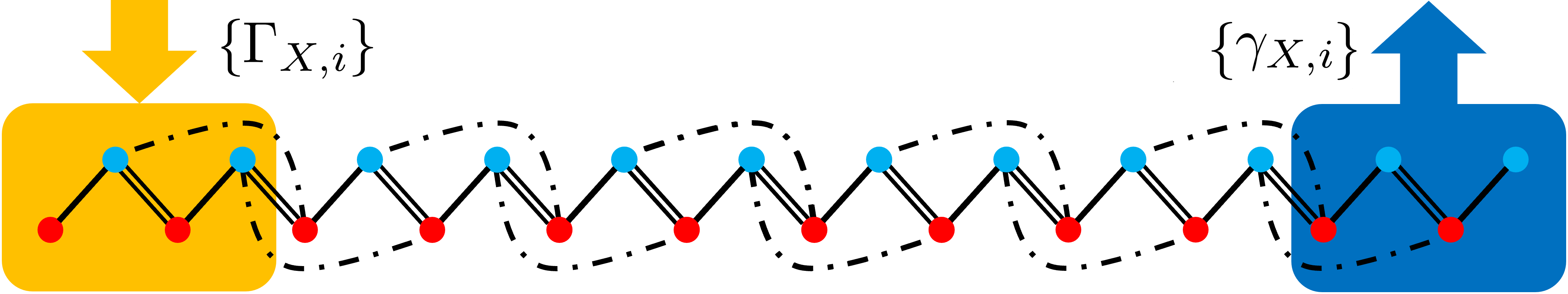}}
\caption{Pictorial representation of the eSSH chain coupled to the bath. The first and last $n$ dimers are coupled with a bath that injects and removes electrons with rates $\Gamma_{X,j}$ ($X=\{A,B\}$, $1\le j \le n$) and $\gamma_{X,j}$ ($X=\{A,B\}$, $N-n \le j \le N$), respectively.}
\label{fig:eSSH_lindblad}
\end{figure}

When the system is at equilibrium, the winding number is one of the standard topological invariants used to characterize the full phase diagram of the eSSH model. It can also be generalized in the presence of chiral symmetry preserving disorder, through the introduction of the DAWN \cite{Mondragon2014,Lin2021}. However, it totally fails in the presence of disorder that breaks the chiral symmetry. To overcome these limitations, in the following we will make use of the EOD topological invariant, recently introduced in Ref.\cite{nava2023_ssh}. In addition to being an experimentally measurable quantity in the out-of-equilibrium regime, it can be effectively employed both in the clean and dirty limits.
%even in presence of chirality breaking disorder. 
\noindent
Following the recipe of Ref.\cite{nava2023_ssh}, we employ the LE formalism to investigate the topological phase in the eSSH model. We induce the system into a NESS, assuming the system coupled to two external thermal baths in the strong bias limit, and we study the time evolution of the system by means of the LE
\begin{equation}
    \dot{\rho}(t)=-i\left[ H,\rho(t)\right]+\sum_k\left(L_k\rho(t)L^\dagger_k-\frac{1}{2}\left\{ L^\dagger_kL_k,\rho(t)\right\}\right)
    \label{eq:Lindblad}
\end{equation}
with $\rho(t)$ the density operator of the system at the time $t$ and $\{L^\dagger_k,L_k\}_k$ a set of operators describing the type of coupling with the bath, called jump operators. In the strong bias limit, we assume that the bath acts as a particle source on the first $n$ dimers of the chain, and as a sink on the last $n$ ones (see Fig.\ref{fig:eSSH_lindblad} for a pictorial representation of the system coupled to the bath). More explicitly, this means that we can parametize the jump operators as
\begin{equation}
    \{L\}_k=\{\{\sqrt{\Gamma_{X,i}}c^\dagger_{X,i}\},\{\sqrt{\gamma_{X,L-i+1}}c_{X,L-i+1}\}\}_{\substack{X=\text{A,B} \\ i=1,\dots, n}}
    \label{eq:bath_operators}
\end{equation}
with the coupling strength $\Gamma_{X,i}$ and $\gamma_{X,i}$ given by:
\begin{align}
    &\Gamma_{X,i}=
    \begin{cases}
        g &\text{ if } i=1,2,\dots,n \text{ and } X=\text{A,B} \\
        0 &\text{ otherwise }
    \end{cases} \\
    &\gamma_{X,i}=
    \begin{cases}
        g &\text{ if } i=L-n+1,\dots,L \text{ and } X=\text{A,B} \\
        0 &\text{ otherwise }
    \end{cases}
    \label{eq:bath_coupling}
\end{align}

\noindent
The EOD is then defined as the average value of the chiral operator
\begin{equation}
    \Bar{\nu}(t)={\rm Tr} [{\Gamma\rho(t)}]=\sum_{i=1}^{N} {\rm Tr}[c_{A,i}^\dagger c_{A,i}  \rho ( t )-c_{B,i}^\dagger c_{B,i}  \rho ( t )]
    \label{eq:eod}
\end{equation}
For a quadratic Hamiltonian, it is possible to write a closed set of equations for the bilinear operators only. Defining the vector $\overrightarrow{c}\equiv( c_{A,1},c_{B,1},\ldots,c_{A,N},c_{B,N}) ^{\top}$, we can write the matrix form system
\begin{equation}
\dot{{\cal C}} ( t )  =i [  {\cal H}^{\top} ( t )   , {\cal C} ( t )  ]+ {\cal G}-
\frac{1}{2} \{  ( {\cal G}+ {\cal R} ), {\cal C} ( t )  \} 
\;\;\;\; , 
\label{eq:HF-master}
\end{equation}
with the bilinear expectation matrix elements  $[ {\cal C} ( t ) ]_{a,b}= {\rm Tr} [c_a^\dagger c_{b}  \rho ( t )]$, the Hamiltonian matrix defined through $H=\overrightarrow{c}^{\dagger} {\cal H} \overrightarrow{c} $ 
and the system-bath coupling matrix elements $ [ {\cal G}]_{a,b}=\delta_{a,b}\sum_{k=1}^{2n}g\delta_{b,k}$ and $ [  {\cal R} ]_{a,b}=\delta_{a,b}\sum_{k=1}^{2n}g\delta_{b,L+1-k}$. The indices $a\equiv (X,i)$ and $b\equiv (Y,j)$ encode both the lattice and dimer labels. 
Under the driving induced by the biased baths, the system evolves in time, asymptotically 
flowing to its unique g-independent NESS, which is determined from the condition $\dot{\rho}=0 \rightarrow \dot{ {\cal C}}(t)=0$. 

By coupling the system to an external bath pumping electron from the left end, and then letting the system evolve to the respective NESS, we are basically populating the zero-energy modes (if there are any) located at the left end of the chain. Since these modes are exponentially localized on the $A$ or $B$ sublattice, each of them gives an integer $\pm1$ contribution to the total EOD. Clearly, to probe topological phases with a winding number higher than one, we need to couple the baths to a number of dimers at least equal to the number of edge states we are interested in detecting.

\noindent
\begin{figure*}
\includegraphics[width=1 \linewidth]{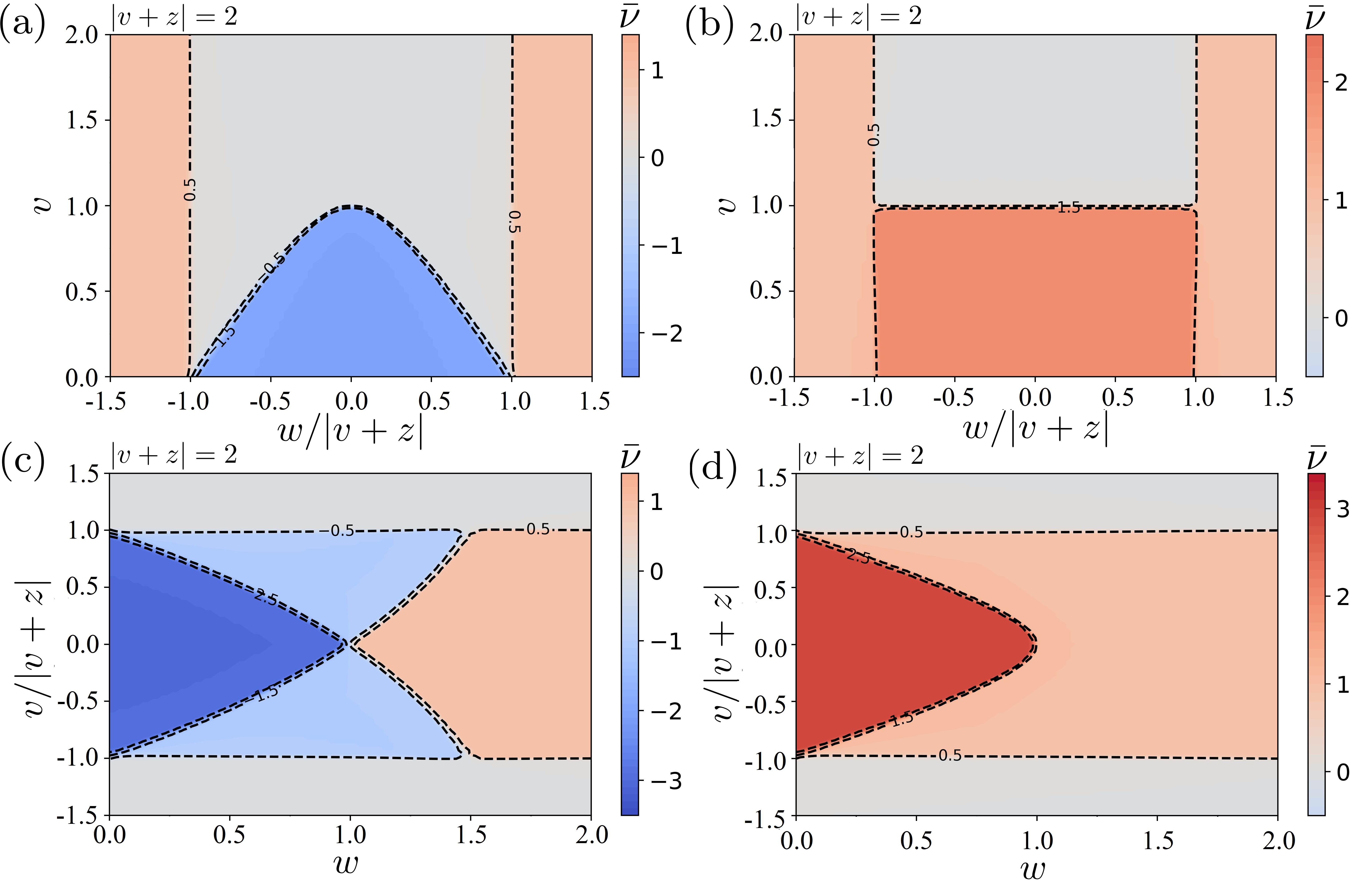}
\caption{Phase diagrams of the eSSH chain for $N=200$ dimers and Hamiltonians a) $H^{\text{A-B}}_2$, b) $H^{\text{B-A}}_2$, c) $H^{\text{A-B}}_3$, and d) $H^{\text{B-A}}_3$. Dashed ones are those at which $\bar{\nu}=n\pm\frac{1}{2}$, with $n\in\mathbb{Z}$.}
\label{fig:eod_clean}
\end{figure*}

\noindent
In Fig.\ref{fig:eod_clean} we report the phase diagrams for Hamiltonian $H^{\text{A-B}}_2$ (panel a), $H^{\text{B-A}}_2$ (panel b), $H^{\text{A-B}}_3$ (panel c), and $H^{\text{B-A}}_3$ (panel d) in the $v-w$ plane with fixed $|v+z|=2$ for a chain of $N=200$ dimers. The EOD perfectly reproduces the result obtained when computing the winding number $\omega$ (see Ref.\cite{Malakar2023} for a comparison). The EOD assumes integer quantized values everywhere in the parameter space, except in proximity to each phase transition with the crossover between phases with different values of the EOD that become sharper with increasing $N$. In Fig.\ref{fig:eod_clean} we plot the isolines along with the EOD assumes semi-integer values to highlight that, already at $N=200$, the crossover region has shrunk significantly.

All the results in this Section apply to the clean limit. In the next Sections, we discuss how phases with different values of the EOD are affected by the presence of different kinds of disorder.

\section{Numerical approach to different realizations of the disorder}
\label{sec:noise}

In real systems, impurities and/or defects may either lead to an enhancement, or to a suppression, of the topogical phase, depending on their specific nature and on their density \cite{Nava2017,nava2023_ssh}. It is, therefore, of the utmost importance to check the robustness of the topological phases of the eSSH in the presence of different kinds of disorder. In the following, we consider several possible realizations of disorder, both uncorrelated, as well as correlated, including the possible breaking of the chiral symmetry of the Hamiltonian.
While the procedure used in this paper is quite general, in the following we focus onto three kinds of disorder:
\begin{enumerate}
    \item \textbf{Chirality preserving disorder with uniform distribution (Type I)} : in this case, each nonzero hopping is independently perturbed (site by site) by adding a different random offset 
        \begin{align}
            v & \rightarrow v_i= v+ \varepsilon_{i,1} \nonumber \\
            w & \rightarrow w_i= w+ \varepsilon_{i,2} \\
            z & \rightarrow z_i= z+ \varepsilon_{i,3} \nonumber
        \end{align}
    where each $\varepsilon_{i,j}$ is drawn from the following uniform probability distribution:
    \begin{equation}
    P[\varepsilon]=
        \begin{cases}
            \frac{1}{2\sqrt{3}W} & \text{ if } -\sqrt{3}W\leq\varepsilon\leq\sqrt{3}W\\
            0 & \text{ otherwise}
        \end{cases}
    \end{equation}
    with zero mean values and standard deviation $W$. As no new hopping term is generated by the disorder, particularly hopping terms that couple sites belonging to the same sublattice, the chiral symmetry is preserved.
    \item \textbf{Correlated chirality preserving disorder with binary distribution (Type II)}: for $n=2$ ($n=3$), we add an offset to the intra(inter)-dimer coupling strength $v$ ($w$), randomly selected between $0$ or $W$ with a binomial distribution:
    \begin{equation}
        P[\varepsilon]=\sigma\delta(\varepsilon)+(1-\sigma)\delta(\varepsilon-W)
    \end{equation}
    where $\sigma$ is the probability for the hopping to remain unperturbed and $W$ the strength of the perturbation. At the same time, the long range hopping is perturbed in such a way that $|v+z|$ ($|w+z|$) is kept constant and equal to $2$. The inter(intra)-dimer hopping is kept constant. More explicitly, the disorder acts on the local hopping as:
    \begin{center}
        \begin{tabular}{c||c}
              $H^{\text{A-B}}_2$ or $H^{\text{B-A}}_2$ & $H^{\text{A-B}}_3$ or $H^{\text{B-A}}_3$ \\
              \hline
             $\begin{aligned}
                 v_i & = v+ \varepsilon_{i} \\
                 w_i & = w \\
                 z_i & = 2-v-\varepsilon_{i}
             \end{aligned}$ &
             $\begin{aligned}
                 v_i & = v \\
                 w_i & = w+\varepsilon_{i} \\
                 z_i & = 2-w-\varepsilon_{i}
             \end{aligned}$ \\
        \end{tabular}       
    \end{center}
    While preserving the chiral symmetry of the Hamiltonian, the Type II disorder can give rise to a finite number of delocalized states in the thermodynamic limit thus allowing for an insulator-to-metal transition \cite{Cheraghchi2005,Izrailev2012}.
    \item \textbf{Correlated chirality breaking disorder with binary distribution (Type III)} : in this case, the Hamiltonian is perturbed by adding a chemical potential term to a random subset of dimers. More explicitly, the perturbation is of the following form:
    \begin{equation}
        \sum_{j}\mu_{j}(c^\dagger_{A,j}c_{A,j}+c^\dagger_{B,j}c_{B,j})
    \end{equation}
    with $\mu_{j}$ coming from the binomial probability distribution:
    \begin{equation}
        P[\mu]=\sigma\delta(\mu)+(1-\sigma)\delta(\mu-W)
        \label{eq:dimer_rand_distr}
    \end{equation}
    with $\sigma$ and $W$ being the perturbation probability and the strength of disorder.
\end{enumerate}

In the following we numerically compute the EOD to spell out the effects of increasing disorder strength on the phase diagram of the eSSH model. We adopt the following recipe:
\begin{enumerate}
    \item For each disorder type and fixed unperturbed values of the hopping strengths $v$, $w$, and $z$, we generate a random disorder configuration, choosing the Hamiltonian parameters through the corresponding probability distribution.
    \item We solve the equation $\dot{ {\cal C}}(t)=0$ to compute the NESS of the perturbed system and the respective EOD, $\bar{\nu}$.
    \item To account for statistical fluctuations, we repeat the procedure over a large amount of disorder realizations $\mathcal{N}$ to compute the disorder averaged EOD
    \begin{equation}
        \langle\bar{\nu}\rangle=\frac{1}{\mathcal{N}}\sum_{r=1}^{\mathcal{N}}\bar{\nu}^{(r)}
    \end{equation}
    (in this paper we set $\mathcal{N}=400$).
    \item To check how much the $\bar{\nu}^{(r)}$'s are peaked around their mean value, we compute the standard deviation $\sigma_{\bar{\nu}} $ of the average EOD, defined as: 
    \begin{equation}
        \sigma_{\bar{\nu}}=\sqrt{\frac{1}{\mathcal{N}}\sum_{r=1}^{\mathcal{N}}\left(\bar{\nu}^{(r)}-\langle\bar{\nu}\rangle\right)^2}
    \end{equation}
    \item Repeating the procedure for all the points in the plane of Fig.\ref{fig:eod_clean} we compute the area associated with each topological phase, i.e., with each value of the EOD, at fixed disorder strength. That is
    \begin{equation}
        \mathcal{A}_{\nu}=\frac{\intop\Theta(\nu+\frac{1}{2}-\left\langle \bar{\nu}\right\rangle )\Theta(\left\langle \bar{\nu}\right\rangle -\nu+\frac{1}{2})dvdw}{\intop dvdw}
    \end{equation}
    with $\Theta(x)$ the Heaviside step function.
\end{enumerate}

Generally speaking, we find that the chirality preserving disorder (Type I and Type II) destroys topological phases in a regular way, in the sense that starting from a non perturbed Hamiltonian with $|\bar{\nu}|>0$, increasing the disorder strength $W$ makes the disorder averaged EOD, $\langle\bar{\nu}\rangle$, approach zero by sequentially assuming all the integer intermediate values. Furthermore, due to the Griffiths effect, a broadening of the transition lines, rather than sharp phase
boundaries, is observed between phases with different disorder averaged EOD \cite{Motrunich2001,Sau2013}. Indeed, near each phase transition,
when averaging over $\mathcal{N}$ different realization of the disorder, the EOD is always quantized for each single realization but some configurations exhibit EOD $\bar{\nu}$ and others $\bar{\nu}\pm1$.  

Conversely, chirality breaking disorder (Type III) gives rise to more interesting outcomes. When the chiral symmetry is weakly perturbed, the eSSH chain still supports edge states in the band gap, and the EOD allows us to detect their presence. When disorder strength increases, the bulk-boundary correspondence is lost and the EOD is not quantized even for a single disorder configuration. However, it is possible to connect the EOD with the spatial distribution of the eigenstates of the system with respect to the sublattices $A$ and $B$.

\section{Strong Disorder Renormalization Group analysis}
\label{sec:SDRG}

Before moving to a full numerical treatment of the LE, in this Section we review the SDRG approach to the eSSH model in order to obtain some hints on the fate of the topological phases in the presence of disorder. The SDRG method, firstly developed for the Heisenberg model in the presence of impurities by Dashgupta, Hu and Ma \cite{Ma1979,Dasgupta1980} and then further developed by Fisher \cite{Fisher1992,Fisher1994,Fisher1995} for the Ising model, is a standard approach to phase transitions in random systems, also in the presence of long range hopping and many body interactions \cite{Huse2023,Igloi2018}.

The SDRG consists of a real space coarse-graining of the Hamiltonian: at each step, a finite-amount of degrees of freedom (spin, boson, fermion, etc.) is integrated over, and all the other couplings are renormalized accordingly. More explicitly, the term in the Hamiltonian having the highest coupling magnitude is diagonalized and a projection onto the corresponding ground state, of the other terms, is performed. It is worth stressing that, as will be evident through this Section, the SDRG approach is suited for finding, at least approximately, the transition line between two different topological phases but gives no hints on the value of the winding number of each topological phase.

Let us consider the most generic chiral symmetric Hamiltonian
\begin{equation}
    \mathcal{H}=\sum_{ij}K_{ij}\left(c^\dagger_{A,i}c_{B,j} + c^\dagger_{B,i}c_{A,j}\right)
    \label{eq:generic_H}
\end{equation}
The SDRG procedure consists of the following steps:
\begin{enumerate}
    \item The strongest hopping, $K_{lm}= \max({\{|K_{ij}|\}})$ with $l<m$, is selected.
    \item The local Hamiltonian depending on $K_{lm}$, i.e.,
        \begin{equation}
            \mathcal{H}_{lm}=K_{lm}(c^\dagger_{A,l}c_{B,m}+c^\dagger_{B,m}c_{A,l})
            \label{eq:localH}
        \end{equation}
    is written as a $4\times 4$ matrix in the occupation number basis $\{\ket{i_{A,l},i_{B,m}}\}$ (if we have more than one strong hopping, Eq.\eqref{eq:localH} can be generalized to include all the terms proportional to them). The eigenvalues and the corresponding eigenstates are given by
    \begin{center}
        \begin{tabular}{l|l}
        Eigenvalue & Eigenstate \\
        \hline
             $E_{1,\pm}=\pm K_{lm}$& $\ket{\psi_{1,\pm}}=\frac{1}{\sqrt{2}}(c^\dagger_l\pm c^\dagger_m)\ket{0,0}$\\
             $E_{0,\pm}=0$            & $\ket{\psi_{0,-}}=\ket{0,0}$; $\ket{\psi_{0,+}}=c^\dagger_lc^\dagger_m\ket{0,0}$
        \end{tabular}
    \end{center}
    \item The global ground state is assumed to be the state $\ket{\psi_{1,-}}$. Since $K_{ij}\le K_{lm}$, we can treat the terms of $\mathcal{H}-\mathcal{H}_{lm}$ depending on $c^\dagger_{A,l}$, $c_{A,l}$, $c^\dagger_{B,m}$, and $c_{B,m}$ perturbatively, using second order perturbation theory.
    \item In this case, the first nonzero contribution is only the second order and is given by
    \begin{equation}
        \sum_{i,\nu} \frac{\bra{\psi_{1,-}}\mathcal{H}-\mathcal{H}_{lm}\ket{\psi_{i,\nu}}\bra{\psi_{i,\nu}}\mathcal{H}-\mathcal{H}_{lm}\ket{\psi_{1,-}}}{E_{-}-E_i}
    \end{equation}
    By neglecting the higher order correction, the net result is that a new effective Hamiltonian will be defined, without the $(A,l)$ and $(B,m)$ degrees of freedom and with the remaining couplings renormalized through the following relation
    \begin{equation}
        \tilde{K}_{ij} =K_{ij}-\frac{K_{il}K_{mj}}{K_{lm}}
        \label{eq:RG_relation}
    \end{equation}
\end{enumerate}

In principle, Eq.\eqref{eq:RG_relation} should be iterated until the renormalized Hamiltonian can be solved by direct diagonalization, or it reduces to a system of which all of its properties are known. 
In general, it is not easy to find a closed form solution for the recursive relation in Eq.\eqref{eq:RG_relation}. However, for some special cases, it is possible to obtain an explicit formula that allows us to analytically detect the phase boundaries. 

Before moving to the eSSH case, let us consider the simpler SSH model in Eq.\eqref{eq:ssh_normal} perturbed by a local disorder that modifies the intra- and inter-dimer hoppings, i.e.,
\begin{equation}
\begin{aligned}
    v_i & =v+\epsilon_{v,i} \\
    w_i & =w+\epsilon_{w,i}
\end{aligned}
\end{equation}
where $\epsilon_{v,i}$ and $\epsilon_{w,i}$ are random numbers coming from some probability distribution $\text{P}[\epsilon]$, that can, in principle, be different for the two hopping strengths.
Upon these assumptions, after $l\gg1$ renormalization steps (see Appendix[\ref{app:SDRG}] for technical details regarding the steps), intra- and inter-dimer hoppings renormalize to
\begin{equation}
\begin{aligned}
    \tilde{v}_i & \approx \text{e}^{l(\left<\ln v\right> - \left<\ln w\right>)}\\
    \tilde{w}_i & \approx \text{e}^{l(\left<\ln w\right> - \left<\ln v\right>)}
\end{aligned}
\label{eq:DDRG_asymp_coupling}
\end{equation}
With $\left<\ln v\right>=\frac{1}{l}\sum_i\ln v_i$ and $\left<\ln w\right>=\frac{1}{l}\sum_i\ln w_i$.

\begin{figure}
\includegraphics[width=\linewidth]{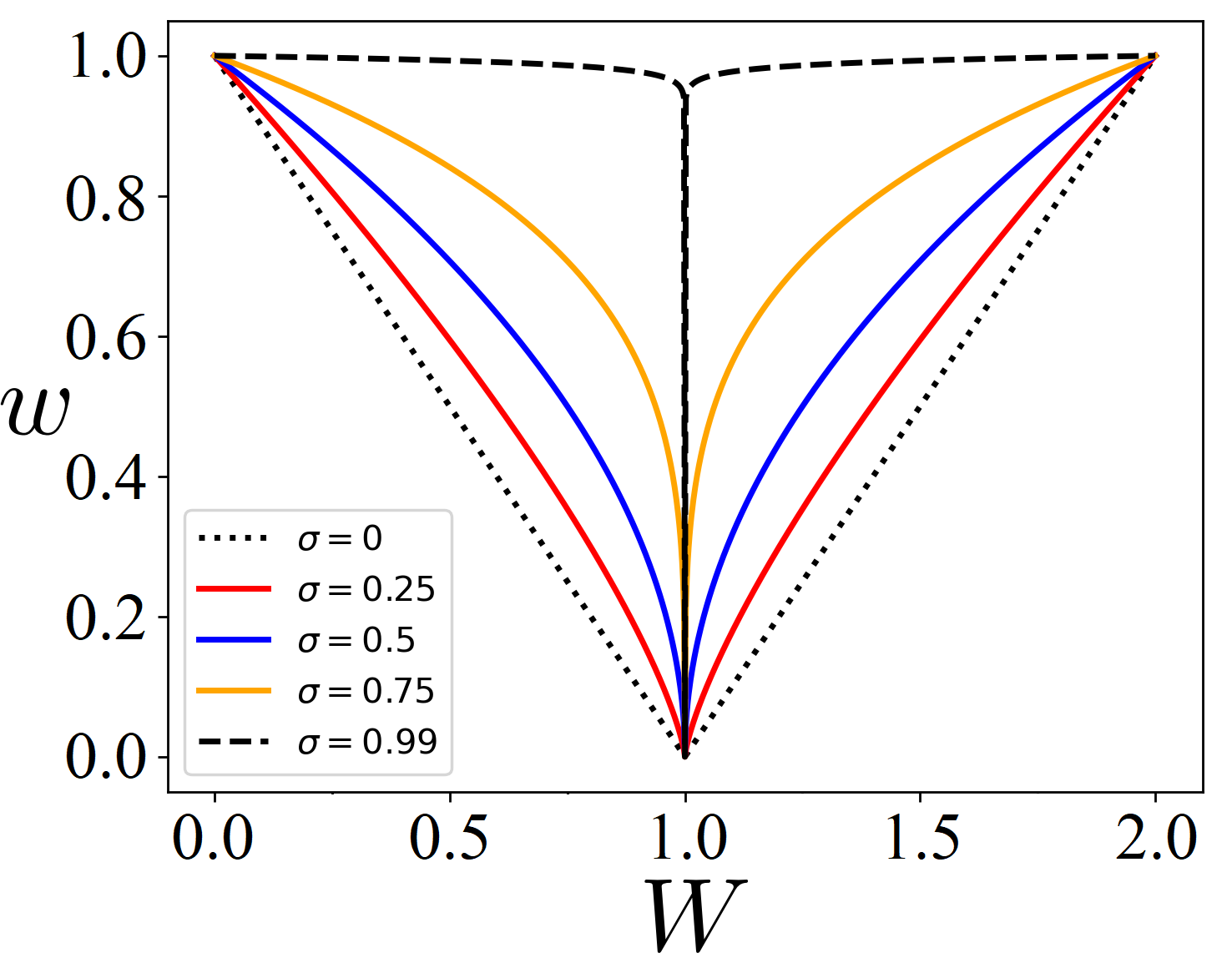}
\caption{Critical line \eqref{eq:v_i_prob_SSH} for different values of $\sigma$ for a SSH chain with random bond disorder.}
\label{fig:w_vs_noise_SSH}
\end{figure}

\noindent
It follows that the SSH chain is in the topological or trivial phase depending on the sign of the exponent: if $\left<\ln v\right> > \left<\ln w\right>$, then all the inter-dimer couplings renormalize to zero, i.e., $\tilde{w}_i\rightarrow 0$, and thus, the chain is in the trivial phase. On the contrary, for $\left<\ln v\right> < \left<\ln w\right>$ all the intra-dimer couplings renormalize to zero, i.e., $\tilde{v}_i\rightarrow 0$, and thus, the chain is in the topological phase. The transition curve, obtained with this RG scheme, is thus given by the condition
\begin{equation}
    \left<\ln v\right> = \left<\ln w\right>
    \label{eq:transition_line}
\end{equation}
Thus, the specific shape depends on the selected probability distribution $\text{P}[\epsilon]$. If the coupling constants can also take negative values, for example when $\epsilon_{v,i}<-v$, we must search for $\max(\{|v_i|\},\{|w_i|\})$, and the transition condition is replaced by
\begin{equation}
    \left<\ln |v|\right> = \left<\ln |w|\right>
    \label{eq:transition_line_abs}
\end{equation}
A simple check of the validity of Eq.\eqref{eq:transition_line_abs} is obtained by looking at the phase diagram of the SSH model in the presence of random bond disorder, studied in Refs.\cite{Liu2022, nava2023_ssh} by means of the DAWN and the EOD, respectively. Assuming all the $w_i$ constant and equal to $w$, while $v$ is a random variable that can assume only two values, with the following binary probability distribution
\begin{equation}
    P[v_i]=\sigma\delta(v_i-1)+(1-\sigma)\delta(v_i-1+W)
    \label{eq:v_i_prob_SSH}
\end{equation}
the critical condition in Eq.\eqref{eq:transition_line_abs} reduces to
\begin{equation}
    w=|1-W|^{1-\sigma}
\end{equation}
\noindent
In Fig.\ref{fig:w_vs_noise_SSH}, we show the transition line as a function of the disorder strength, $W$, and of the inter-dimer hopping, $w$, for different values of the disorder probability $\sigma$. The results are in perfect agreement with panel a) of Fig.1 and panel a) and b) of Fig.3 of Ref.\cite{Liu2022}, as well as with Fig.11 of Ref.\cite{nava2023_ssh}. 

We can now employ the SDRG approach to the eSSH in the presence of disorder. Even though it is not possible to analytically solve Eq.\eqref{eq:RG_relation} in the presence of all three hopping $v$, $w$, and $z$, we can find a closed form solution along the special cases shown in Fig.\ref{fig:SSH_extreme}, i.e., when one of the hoppings is tuned to zero. To do so, we first promote the two nonzero hoppings to local variables and then, since the system decouples into distinct SSH chains, we can find the transition line using Eq.\eqref{eq:transition_line}. It is worth stressing out that the disorder can generally perturb all the hoppings, including the one we have assumed to be zero in the unperturbed case. As a consequence, the eSSH chain is not truly decoupled into distinct SSH chains, and the SDRG transition line discussed in the following is only valid in the limit of very weak disorder. 
\begin{figure}
\includegraphics[width=\linewidth]{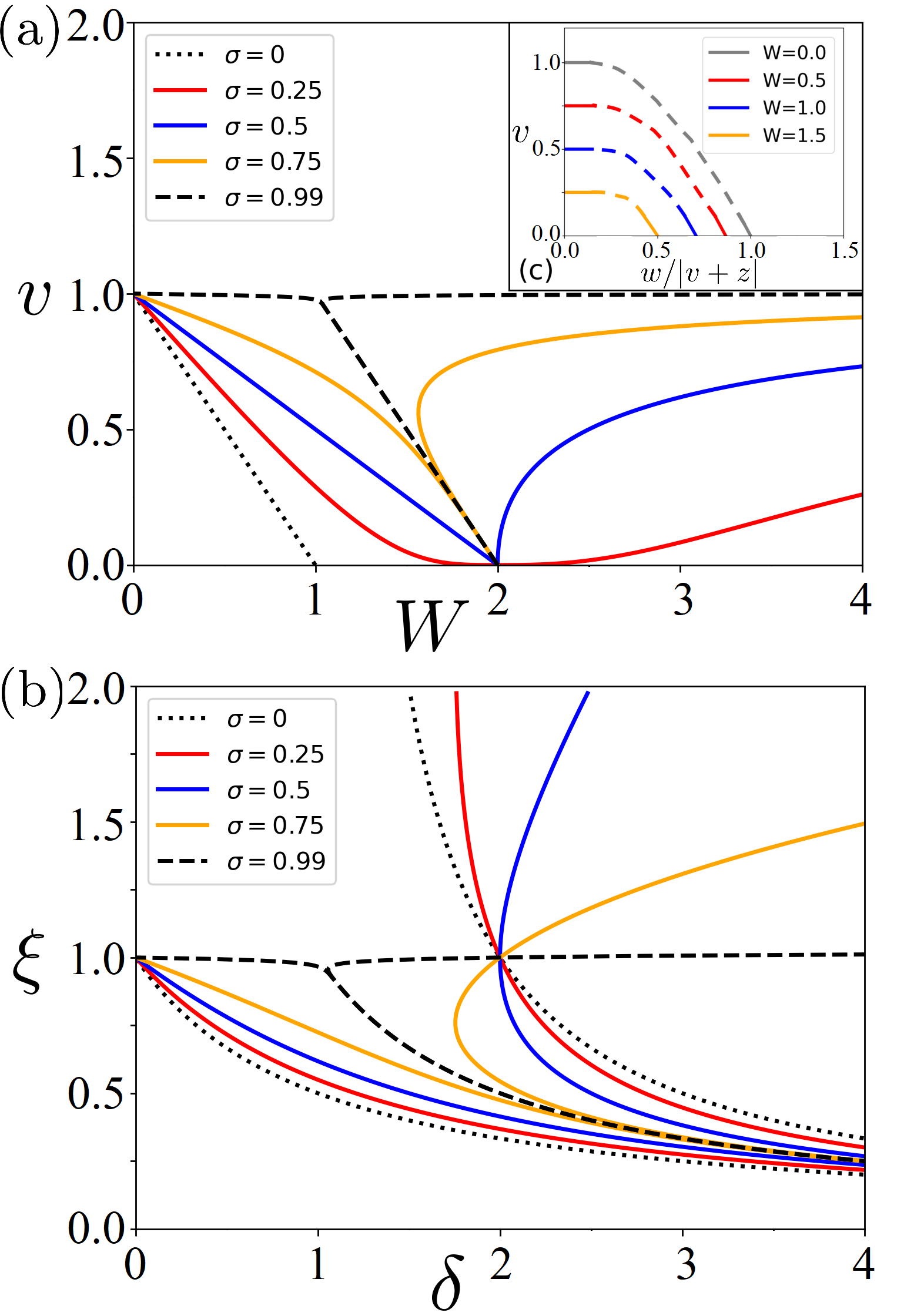}
\caption{Critical line for Type II disorder along the extreme cases obtained by setting: panel a) $w=0$, as a function of $v$ and $W$ for different values of $\sigma$; panel b) $v=0$ as a function of $\xi=\frac{w}{2-v}$ and $\delta=\frac{W}{w}$ for different values of $\sigma$.
The inset of panel a) shows the putative phase diagram, in the $v-w$ plane, obtained combining the results in panel a) and b) as a function of disorder strength $W$ and at fixed $\sigma=0.5$.}
\label{fig:transition_lines_extremes}
\end{figure}

\noindent
Solving Eq.\eqref{eq:RG_relation} assuming a zero inter-dimer hopping, we obtain
\begin{center}
    \begin{tabular}{c||c}
            Disorder     &  $\left<\ln v\right> = \left<\ln z\right> \quad (w=0)$\\
            \hline
            Type I &  $v=z=1\quad$ \\
            \hline
            Type II &  
            $W_{c,\pm}=\frac{2}{1\pm\left(\frac{v}{2-v}\right)^\frac{\sigma}{1-\sigma}}-v$
    \end{tabular}    
\end{center}
while for zero intra-dimer hopping we have
\begin{center}
    \begin{tabular}{c||c}
            Disorder     & $\left<\ln w\right> = \left<\ln z\right> \quad (v=0)$\\
            \hline
            Type I & $w=2-v\quad $ \\
            \hline
            Type II & 
            $W_{c,\pm}=\frac{1\pm\xi^{\frac{1}{1-\sigma}}}{\xi}w,\text{  } \xi=\frac{w}{2-v}$
    \end{tabular}    
\end{center}
The non trivial transition line for Type II disorder is shown in Fig.\ref{fig:transition_lines_extremes} for $w=0$ (panel a) and $v=0$ (panel b).
All the results above are true at very weak disorder strength. However, the critical values for Type II disorder with $\left<\ln v\right> = \left<\ln z\right>$ is exact as the inter-dimer term is not affected by this kind of disorder. Regarding the other ones, the agreement is less precise, since the relation relies on the assumption that the zero hopping remains unperturbed even in the presence of disorder, which does not hold in general.
Looking at the sketch of the putative transition lines shown in the inset of Fig.\ref{fig:transition_lines_extremes}, and obtained combining the results of panel a) and b), even at weak disorder strength, the area below each curve corresponding to the topological phase with $w=\pm2$ (depending on whether we are considering $H_2^{B-A}$ or $H_2^{A-B}$) shrinks along the vertical axis. Conversely, for Type I disorder (looking at the table above) one can expect that the transition point is untouched by a weak amount of disorder.

Summarizing, while the SDRG represents a good tool to probe the topological phase transition in the parameter space, it suffers some limitations.
In general, closed formulas for the hopping strengths are not available. As a consequence, one should rely on numerical results. At the same time, it cannot give any information about the winding number of each topological phase, nor the width of the Griffiths phase associated at each transition line.
For these reasons, in the following we will implement the EOD method, which does not suffer the limitations highlighted above. 

\section{EOD numerical results}
\label{sec:numerical_results}

In this Section, we compute the disorder averaged EOD, $\langle\bar{\nu}\rangle$, together with its standard deviation, $\sigma_{\bar{\nu}}$ and the area occupied by each topological phase, $\mathcal{A}_{\nu}$.

\subsection{Chirality preserving disorder}
\label{subsec:chiral}

\begin{figure}
\includegraphics[width=\linewidth]{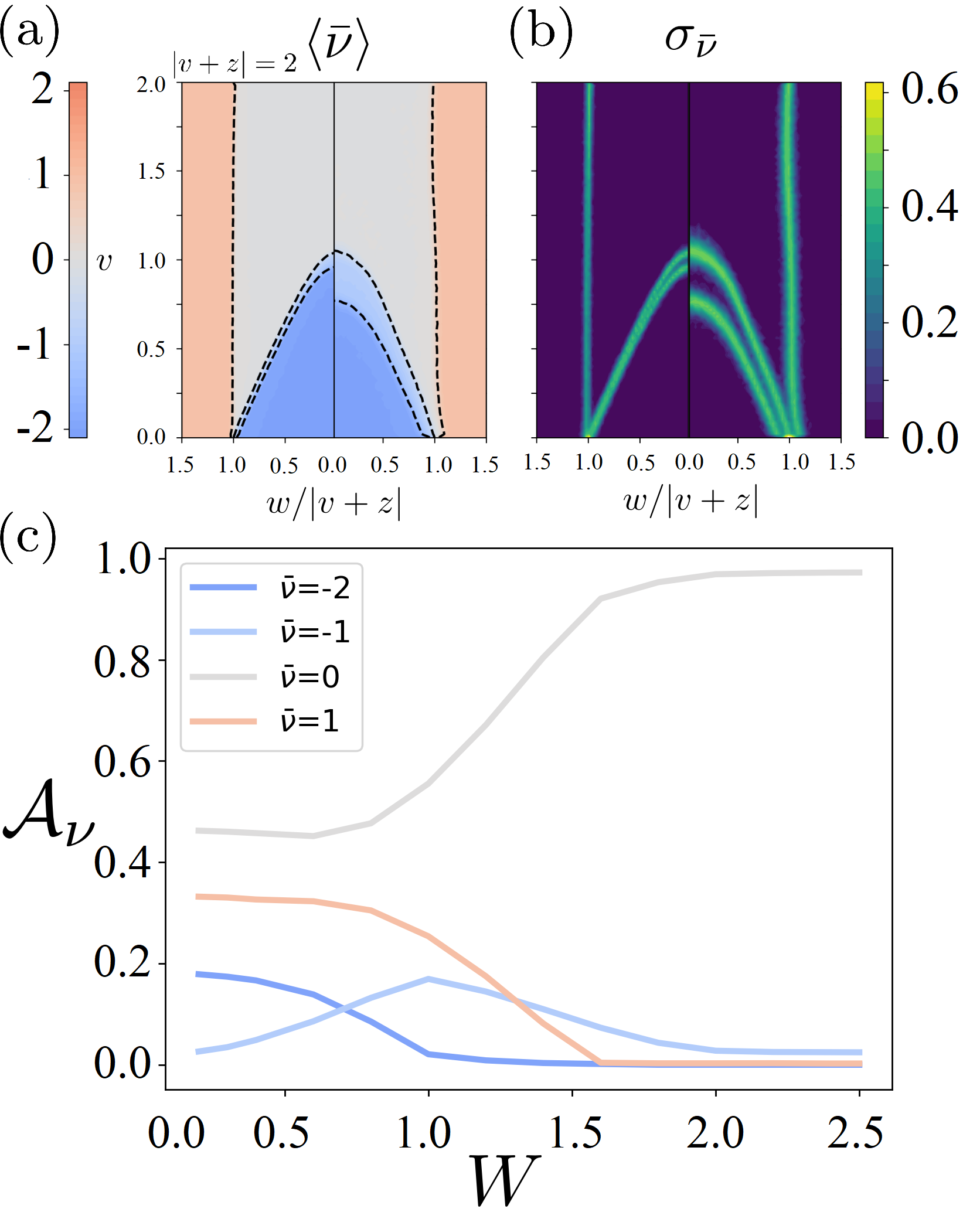}
\caption{Phase diagrams of a $H_2^{A-B}$ eSSH model of length $L=500$ and Type I disorder with $\sigma=0.5$ averaged over $\mathcal{N}=400$ disordered realizations. Panel a) $\langle\bar{\nu}\rangle$ for disorder strength $W=0.3$ (left hand side) and $W=0.6$ (right hand side). Panel b) $\sigma_{\bar{\nu}}$ computed with the same parameter of panel a). Panel c) $\mathcal{A}_{\nu}$ for each topological phase as a function of $W$.}
\label{fig:AB2_typeI}
\end{figure}

Let us start with the $H_2^{A-B}$ eSSH model in the presence of Type I disorder. The main results are shown in Fig.\ref{fig:AB2_typeI} (to be compared with panel a) of Fig.\ref{fig:eod_clean}). Since the phase diagram is symmetrical with respect to a change of sign of the hopping strength, only positive values are considered. In panel a) of 
Fig.\ref{fig:AB2_typeI} the disorder averaged EOD is plotted for two different values of the disorder strength: $W=0.3$ (left hand side) and $W=0.6$ (right hand side). Two main behaviors emerge. The first one is a "buffer region", with $\langle\bar{\nu}\rangle=-1$, between the two phases with $\langle\bar{\nu}\rangle=0$ and $\langle\bar{\nu}\rangle=-2$. The size of this region, which is absent in the clean case, starts to grow with disorder,
engulfing part of the parameter space occupied by the phase with $\langle\bar{\nu}\rangle=-2$ and, at the same time, limiting the development of the phase with $\langle\bar{\nu}\rangle=0$. Indeed, looking at the left and right hand sides of panel a) of Fig.\ref{fig:AB2_typeI} we can observe how the top transition line of the $\langle\bar{\nu}\rangle=-1$ phase is stable up to $W=0.6$, while the bottom transition line moves significantly downwards. Other important information is recovered looking at the standard deviation of the EOD, as reported in panel b) of the same figure. It is clear that each transition line is not sharp, as in the clean case, but exhibits a finite width, which is the signature of a Griffiths phase transition \cite{Motrunich2001,Sau2013}. However, the emerging topological phase with $\langle\bar{\nu}\rangle=-1$ has a strong bulk region, where the standard deviation is strictly zero, between the two finite width transition regions. This is an evidence that this region is not an effect of a statistical combination of an equal number of configurations in the $\langle\bar{\nu}\rangle=0$ trivial phase and the $\langle\bar{\nu}\rangle=-2$ topological phase but rather a truly, robust disorder induced, $\langle\bar{\nu}\rangle=-1$ topological phase. 

A hint about the nature of this phase can be recovered along the line with $w=0$. In this case, as shown in panel c) of Fig.\ref{fig:SSH_extreme}, the eSSH chain is decoupled into two simple SSH models. Deep inside the $\langle\bar{\nu}\rangle=-1$ phase, where $\sigma_{\bar{\nu}}=0$, the effect of the disorder is to induce a topological-to-trivial transition exactly on one of the two chains, while the other one remains protected. While the SDRG approach can describe the main behavior of the lower transition line, i.e., the retreating of the topological phase with the higher winding number, as shown in the inset of Fig.\ref{fig:transition_lines_extremes}, it cannot catch the existence of the buffer region because, in the SDRG approach, the two chains in which the original models decouple at $w=0$ are treated as uncorrelated, so both of them are always assumed in the same phase. Finally, panel c) of Fig.\ref{fig:AB2_typeI} gives us a global picture of the fate of each topological region with increasing disorder strength $W$. The topological region characterized by $\langle\bar{\nu}\rangle=1$ is robust to disorder up to $W\approx1$, after which it begins to be absorbed into the trivial phase. At the same time, the topological region with $\langle\bar{\nu}\rangle=-2$ has been replaced by the buffer one that, in turn, is replaced by the trivial phase at stronger values of $W$. In summary, a sort of hierarchy is observed. The topological phase with the higher value of the winding number is the first to be destroyed until, at strong values of the disorder, the trivial phase is the sole survivor. 

\begin{figure}
\includegraphics[width=\linewidth]{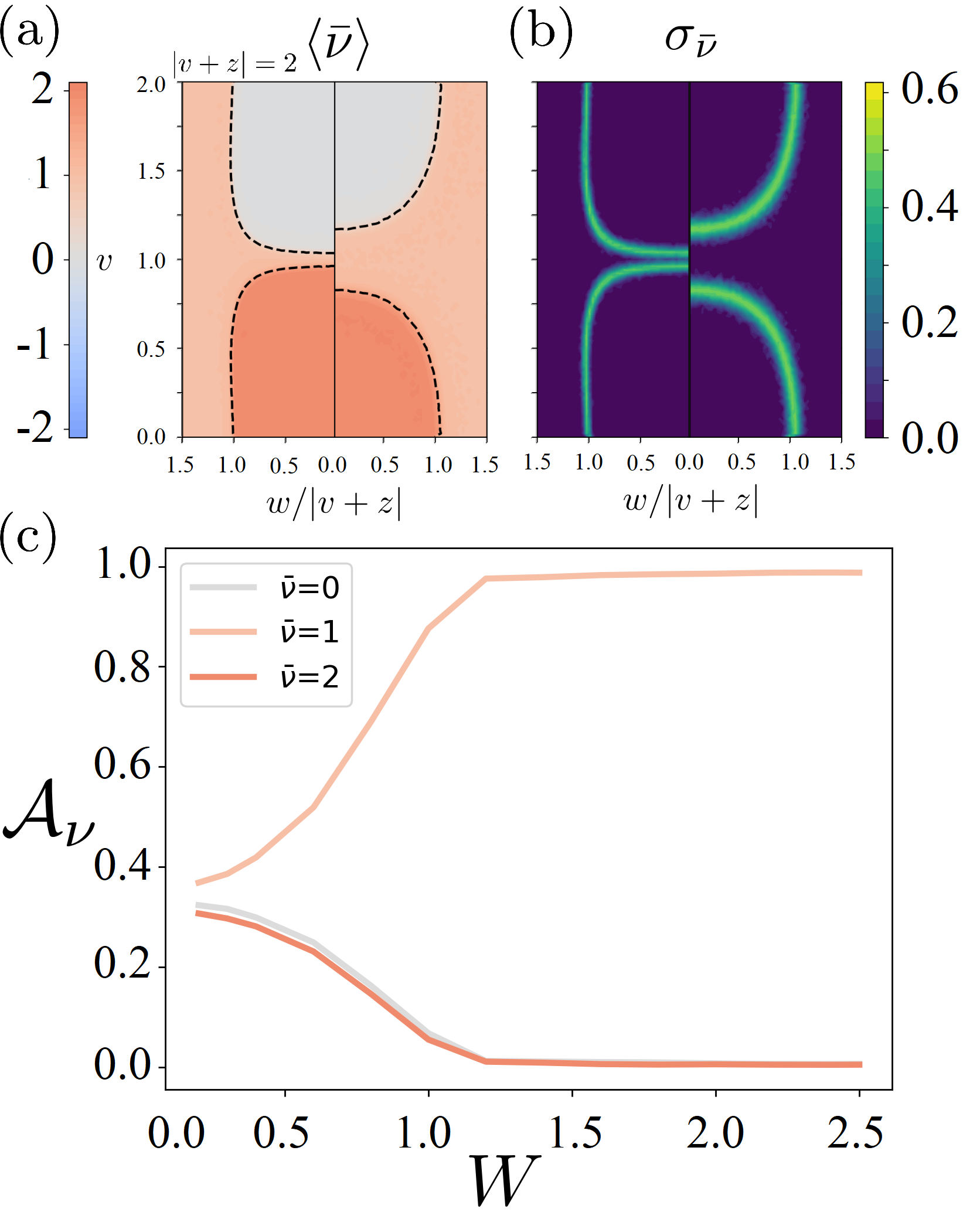}
\caption{Phase diagrams of a $H_2^{B-A}$ eSSH model of length $L=500$ and Type I disorder with $\sigma=0.5$ averaged over $\mathcal{N}=400$ disordered realizations. Panel a) $\langle\bar{\nu}\rangle$ for disorder strength $W=0.3$ (left hand side) and $W=0.6$ (right hand side). Panel b) $\sigma_{\bar{\nu}}$ computed with the same parameter of panel a). Panel c) $\mathcal{A}_{\nu}$ for each topological phase as a function of $W$.}
\label{fig:BA2_typeI}
\end{figure}

Let us now discuss the effect of Type I disorder on the $H_2^{B-A}$ eSSH model, which in the clean limit exhibits phases with positive winding number only (as shown in panel b of Fig.\ref{fig:eod_clean}). Looking at panel a) and b) of Fig.\ref{fig:BA2_typeI}, a buffer region with $\langle\bar{\nu}\rangle=1$, separating the trivial and the $\langle\bar{\nu}\rangle=2$ non trivial phases, emerges. This region merges consistently with the $\langle\bar{\nu}\rangle=1$ region located at $|w|>2$, and already present in the clean case. The buffer region is again well defined, as highlighted by a strictly zero standard deviation, and grows in site with the disorder strength $W$ until, at $W\approx1.2$, it dominates the whole phase diagram, as shown in panel c) of Fig.\ref{fig:eod_clean}. Again, a hierarchy is observed with regions characterized by high values of the winding number suppressed in favor of phases with winding numbers zero or one. Compared with the previous case, in the $B-A$ model the $\langle\bar{\nu}\rangle=1$ is more robust than the zero one because the topological edge state is truly on the last side of the chain while in the $A-B$ case it lies on the $B$ sublattice. However, it is worth noting that, up to disorder strength of $W\approx0.5$, the nontrivial phase with $\langle\bar{\nu}\rangle=2$ is mainly preserved.

\begin{figure}
\includegraphics[width=\linewidth]{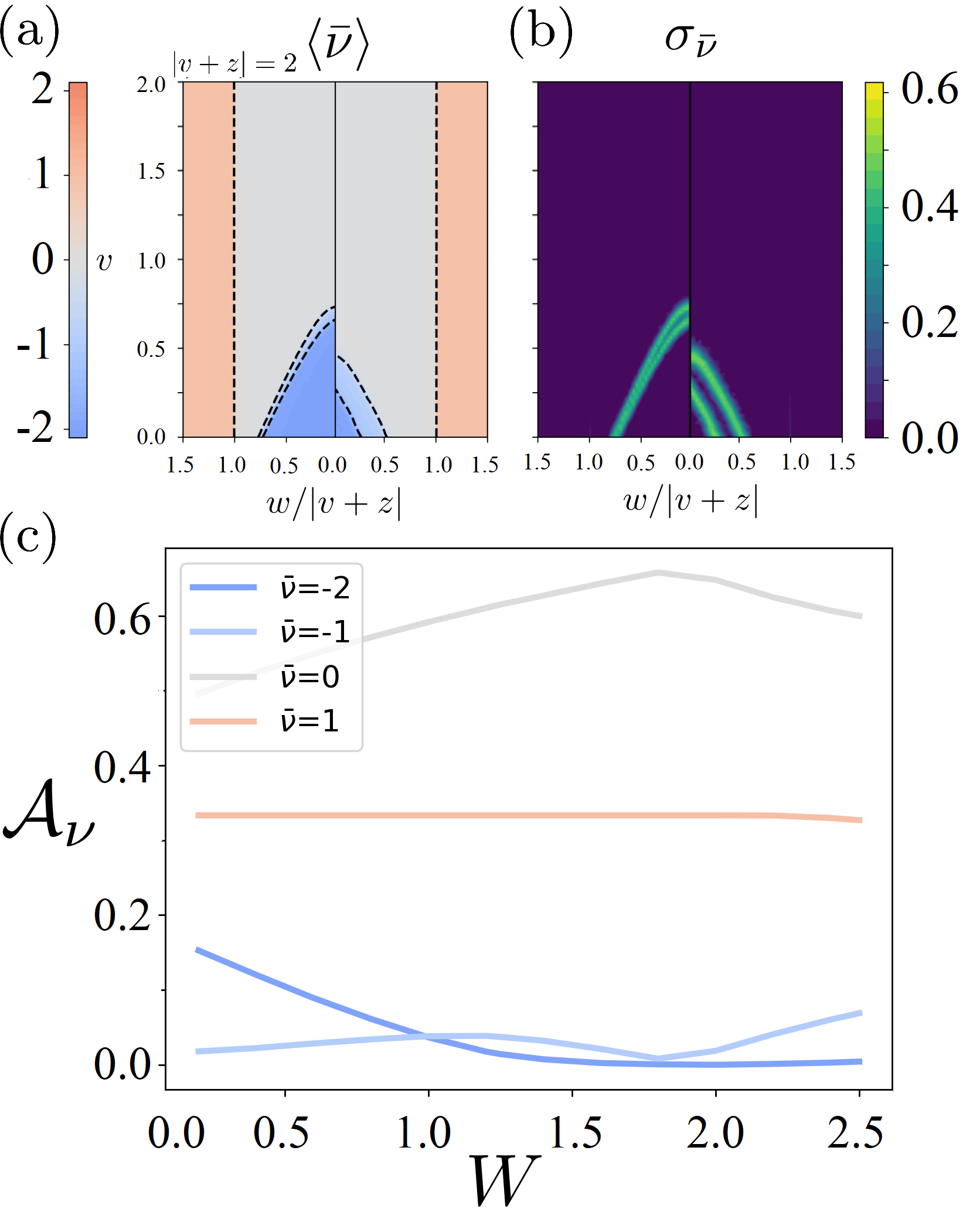}
\caption{Phase diagrams of a $H_2^{A-B}$ eSSH model of length $L=500$ and Type II disorder with $\sigma=0.5$ averaged over $\mathcal{N}=400$ disordered realizations. Panel a) $\langle\bar{\nu}\rangle$ for disorder strength $W=0.6$ (left hand side) and $W=1.25$ (right hand side). Panel b) $\sigma_{\bar{\nu}}$ computed with the same parameter of panel a). Panel c) $\mathcal{A}_{\nu}$ for each topological phase as a function of $W$.}
\label{fig:AB2_typeII}
\end{figure}

In Fig.\ref{fig:AB2_typeII} $\langle\bar{\nu}\rangle$, $\sigma_{\bar{\nu}}$ and $\mathcal{A}_\nu$ are shown for the $H_2^{A-B}$ eSSH in the presence of Type II disorder. Since that Type II disorder does not act on the inter-dimer hopping, the vertical transition line between the trivial and the $\langle\bar{\nu}\rangle=1$ phase is not affected. On the other hand, as in the Type I counterpart, a buffer phase emerges but it is less robust than the previous case with both the $\langle\bar{\nu}\rangle=-1$ and $\langle\bar{\nu}\rangle=-2$ phases retreating in favor of the trivial one. However, at strong disorder values, i.e., $W\approx1.8$, disorder leads to a reentrant topological phase with $\langle\bar{\nu}\rangle=-1$ as highlighted in panel c) of Fig.\ref{fig:AB2_typeII}.

\begin{figure}
\includegraphics[width=\linewidth]{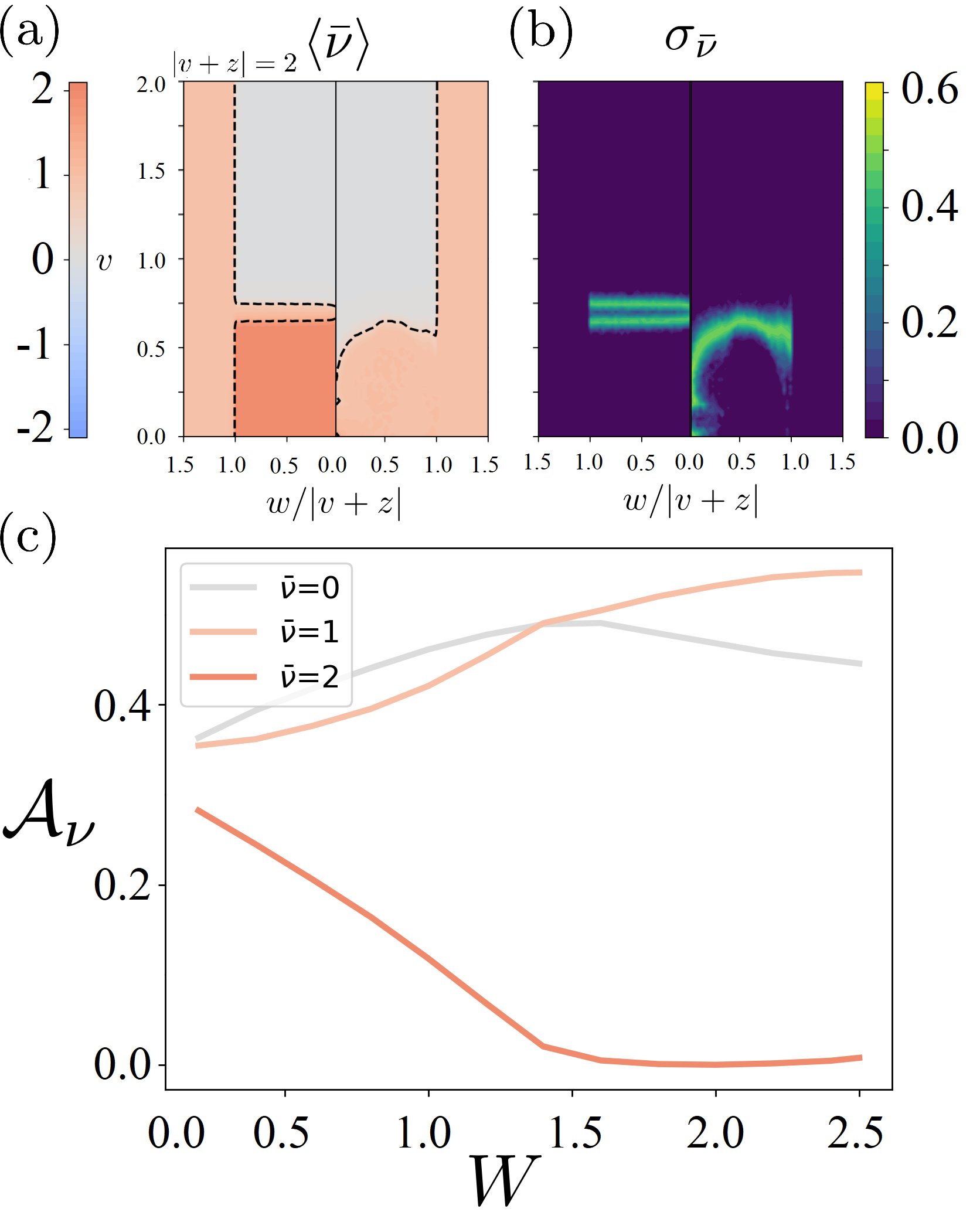}
\caption{Phase diagrams of a $H_2^{B-A}$ eSSH model of length $L=500$ and Type II disorder with $\sigma=0.5$ averaged over $\mathcal{N}=400$ disordered realizations. Panel a) $\langle\bar{\nu}\rangle$ for disorder strength $W=0.6$ (left hand side) and $W=1.25$ (right hand side). Panel b) $\sigma_{\bar{\nu}}$ computed with the same parameter of panel a). Panel c) $\mathcal{A}_{\nu}$ for each topological phase as a function of $W$.}
\label{fig:BA2_typeII}
\end{figure}

\noindent
A similar behavior is shared by $H_2^{B-A}$ eSSH and Type II disorder, as shown in Fig.\ref{fig:BA2_typeII}. Increasing $W$, the $\langle\bar{\nu}\rangle=2$ phase gives up its place to the $\langle\bar{\nu}\rangle=1$ one in its turn embedded by the trivial one. However, at $W\approx1.5$, disorder enhances the topological phase with $\langle\bar{\nu}\rangle=1$, reversing the previous trend. The $\langle\bar{\nu}\rangle=2$ phase is totally suppressed at $W\approx1.5$. Indeed, as shown in the right hand side of panel b), only one of the two Griffiths transition lines survives at strong disorder, with the other one abruptly pushed down.

\begin{figure}
\includegraphics[width=0.8 \linewidth]{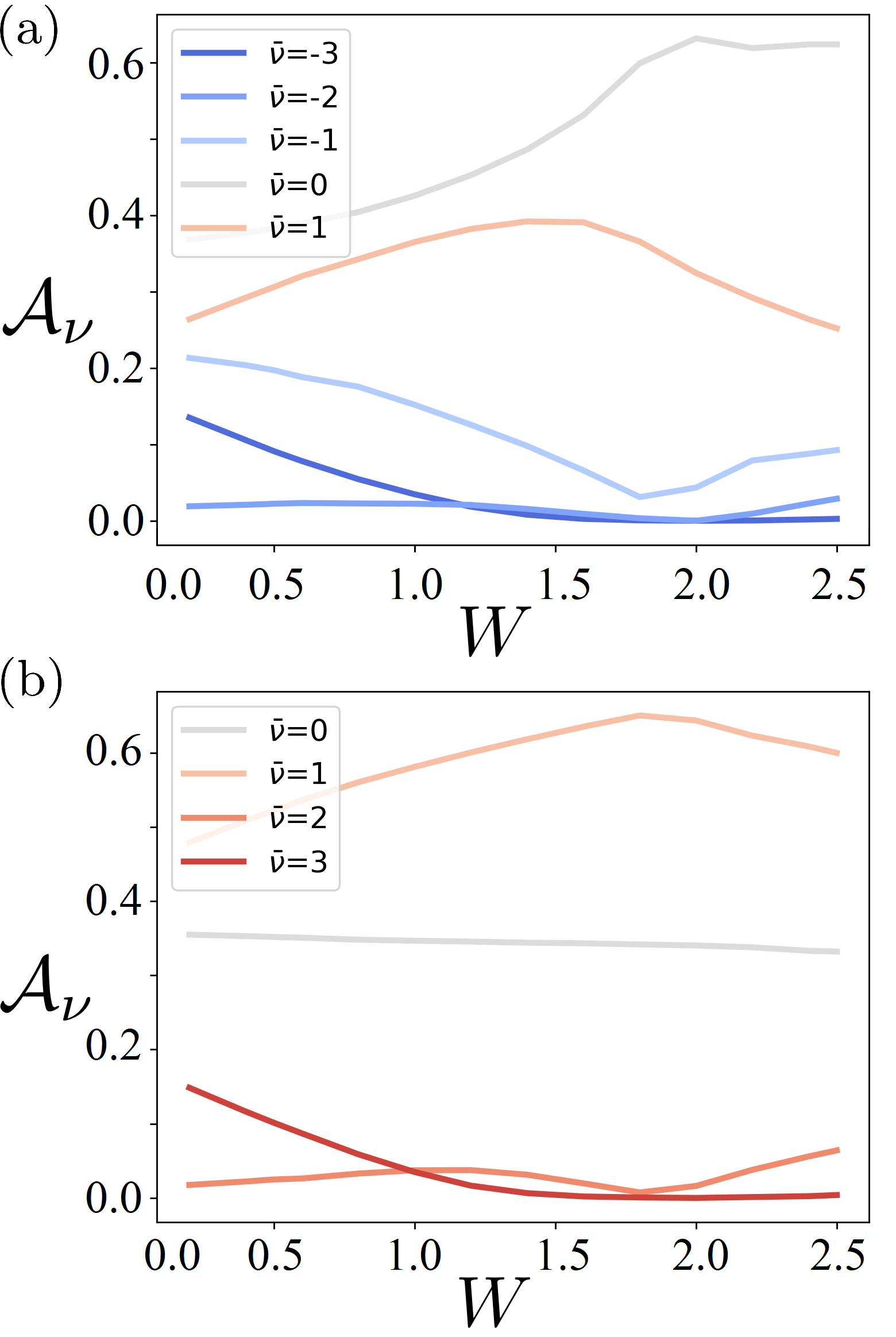}
\caption{$\mathcal{A}_{\nu}$ for each topological phase as a function of $W$ for eSSH model described by $H_3^{A-B}$ (panel a) and $H_3^{B-A}$ (panel b).}
\label{fig:n3_typeII}
\end{figure}

Finally, the hierarchical disappearance of the phases with higher EOD, the appearance of buffer phases with intermediate values of the EOD, and the existence of a reentrant disorder induced topological phase with EOD=1, are preserved with increasing $n$, as shown for example in Fig.\ref{fig:n3_typeII} for n=3 and Type II disorder. In panel a) we observe, in the case of $H_3^{A-B}$, a quick suppression of the $\langle\bar{\nu}\rangle=-3$ phase, followed by a slower suppression of the $\langle\bar{\nu}\rangle=-1$ phase that resists to stronger values of the disorder. At the same time, the $\langle\bar{\nu}\rangle=-2$ phase remains squeezed between the reentrant $\langle \bar{\nu}\rangle=-3$ phase and the enlarging $\langle \bar{\nu}\rangle=-1$ one.  For weak disorder strength we observe the trivial phase slowly replacing the negative topological phase, even though at strong disorder a reentrant topological phase is observed for the negative EOD phases, at the expense of the positive one. Similarly, for $H_3^{A-B}$ shown in panel b), the hierarchy $\langle\bar{\nu}\rangle=3 \rightarrow \langle\bar{\nu}\rangle=1$ is observed with a $\langle\bar{\nu}\rangle=2$ buffer phase in between, that is enhanced at strong $W$. 

\subsection{Chirality breaking disorder}
\label{subsec:nonchiral}

\begin{figure}
\includegraphics[width=\linewidth]{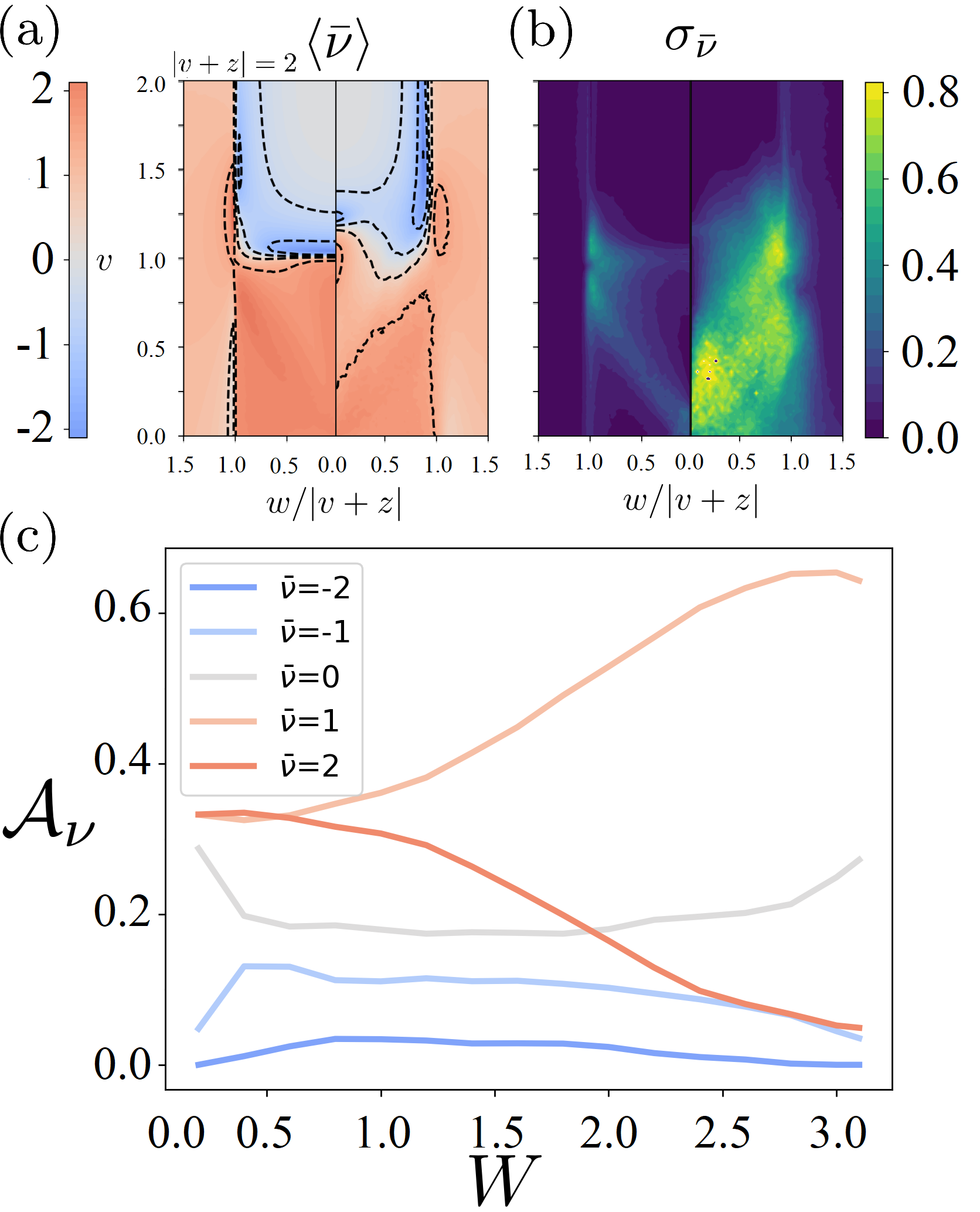}
\caption{Phase diagrams of a $H_2^{B-A}$ eSSH model of length $L=500$ and Type III disorder with $\sigma=0.5$ averaged over $\mathcal{N}=400$ disordered realizations. Panel a) $\langle\bar{\nu}\rangle$ for disorder strength $W=0.5$ (left hand side) and $W=1.8$ (right hand side). Panel b) $\sigma_{\bar{\nu}}$ computed with the same parameter of panel a). Panel c) $\mathcal{A}_{\nu}$ for each topological phase as a function of $W$.}
\label{fig:BA2_typeIII}
\end{figure}

In the presence of chirality breaking disorder the Hamiltonian no longer anticommutes with the chiral operator and eigenstates are no longer expected to appear in pairs, symmetric with respect to the zero value. Valence and conduction bands are expected to merge with each other, and the bulk-boundary correspondence is lost \cite{Gonzalez2019,nava2023_ssh}. In general, gapped configurations hosting localized edge states still exist even in the absence of the chiral symmetry, if the disorder is not too strong. However, they are totally washed out at higher values of the disorder strength. Looking at the left hand side of panels a) and b) of Fig.\ref{fig:BA2_typeIII} we see extended regions in the phase diagram characterized by an integer value of $\langle\bar{\nu}\rangle$ and zero standard deviations. These regions correspond to a disorder resilient topological phase, with localized edge states even with broken chiral symmetry. However, as disorder strength increases (see the right hand side of panels a) and b)), $\langle\bar{\nu}\rangle$ is no longer quantized across the entire parameter space, and its standard deviation is heavily different from zero everywhere. Furthermore, while the phases with $\langle\bar{\nu}\rangle=2$ tend to be destroyed in favor of the $\langle\bar{\nu}\rangle=1$ phase, just as in the presence of chirality preserving disorder, an unexpected region with $\langle\bar{\nu}\rangle=-1$ appears inside the nontrivial region, before shrinking at high values of $W$.

\begin{figure}
\includegraphics[width=\linewidth]{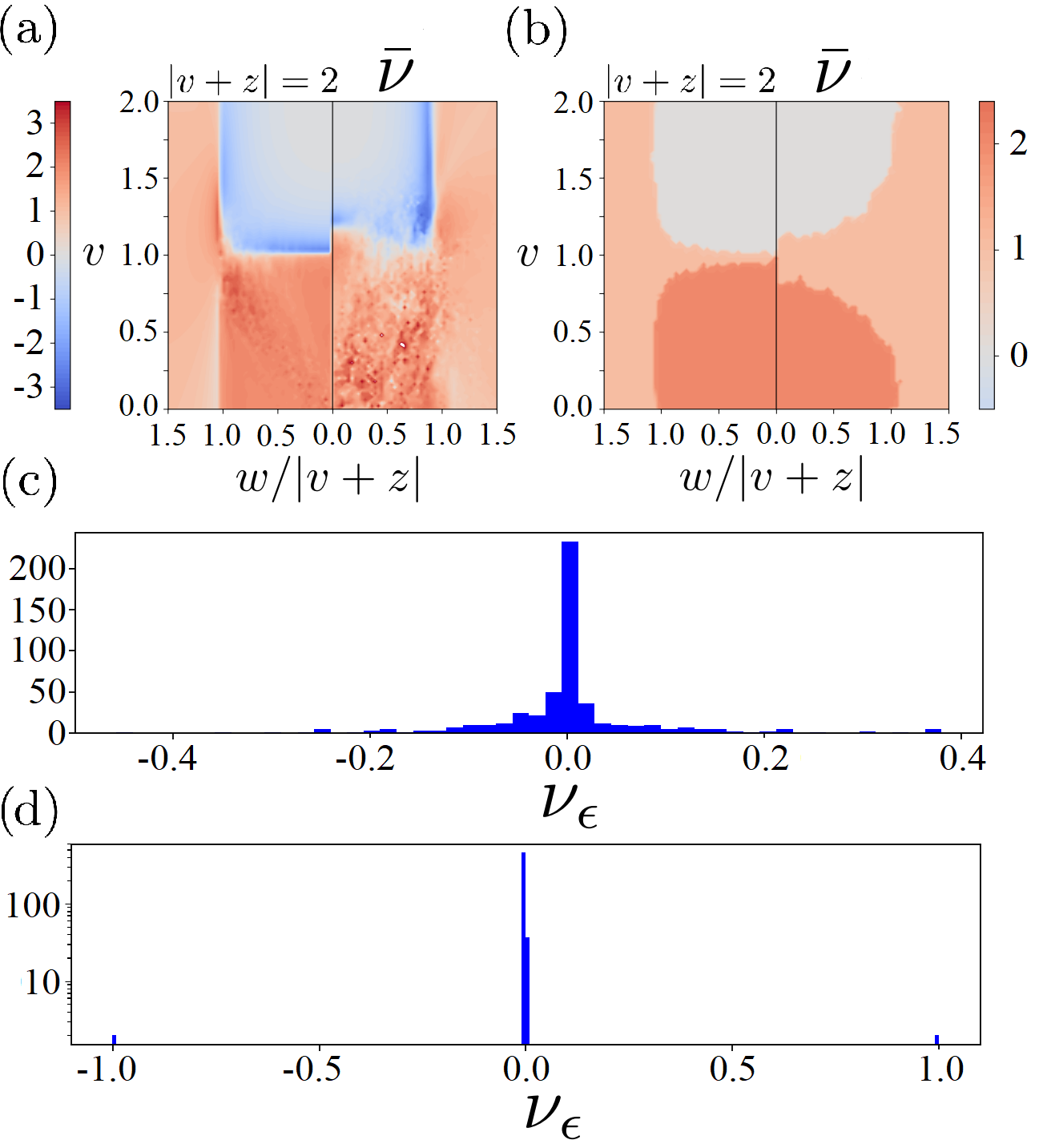}
\caption{Panel a) EOD for a single Type III disorder configuration of a $H_2^{B-A}$ eSSH model of length $L=500$ with $\sigma=0.5$ and $W=0.5$ (left hand side) and $W=1.8$ (right hand side). Panel b) same as a) but for Type I disorder with $W=0.3$ (left hand side) and $W=0.6$ (right hand side). Panel c) histogram of the distribution of the eigensystem EOD for $v=1.25$, $w=1.7$ and $W=1.8$ for Type III disorder. Panel d) same as c) but with $v=0.25$, $w=0.5$, and $W=0.6$ for Type I disorder.}
\label{fig:BA2_typeIII_single}
\end{figure}

The region of the parameter space characterized by nonzero standard deviation and unquantized $\langle\bar{\nu}\rangle$ tends to cover the full phase diagram at strong values of the disorder strength.
However, this phase is significantly different from the true Griffiths phase observed in the presence of chirality preserving disorder of Types I and II. In panels a) and b) of Fig.\ref{fig:BA2_typeIII_single} we show the EOD of an eSSH described by $H_2^{B-A}$ for a single disorder configuration of Type I or Type III (as extracted from panel a) of Fig.\ref{fig:BA2_typeI} and Fig.\ref{fig:BA2_typeIII} respectively). At both weak and strong disorder, the EOD of each chirality preserving disordered configuration is quantized along the full parameter space, and the different topological phases are separated by a jagged but sharp transition line. The transition line smoothly changes for each configuration and the transition region estimated by averaging over a huge number of configurations gives rise to the Griffiths phase, characterized by non zero standard deviation of the EOD over a small region of the parameter space. On the contrary, in the presence of chirality breaking disorder, the EOD of each single disorder configuration is not quantized over a wide region of the parameter space, with this effect more and more evident at increasing $W$. This effective nonchiral camouflaged Griffiths phase (NCCG) is thus the effect of an average over a huge number of configurations each of which has unquantized EOD. Clearly, looking at both $\sigma_{\bar{\nu}}$ and at the EOD allows us to distinguish the NCCG phase from the true Griffiths one. It remains to understand whether, in the NCCG phase, some topology is still present and eventually how it is related to the (disorder averaged) EOD. 
Introducing the Hamiltonian eigenvector $\gamma^\dagger_\epsilon=\sum_j\psi_{\epsilon,j} c^\dagger_j$ and neglecting the correlations between different eigenvectors, the EOD can be written as
\begin{equation}
    \bar{\nu} \approx \sum_{\epsilon}^L \theta_{\epsilon} \nu_{\epsilon}
\end{equation}
where $\nu_{\epsilon}$ is the (equilibrium) EOD of a single Hamiltonian eigenstate
\begin{equation}
    \nu_{\epsilon} = \sum_{j=1}^L(-1)^{j+1}\left|\psi_{\epsilon,j}\right|^2
\end{equation}
and $\theta_{\epsilon}$ its occupation probability, given by
\begin{equation}
    \theta_{\epsilon}=\sum_{j,j'}\psi_{\epsilon,j}\theta_{j,j'}\psi^*_{\epsilon,j'}
    \label{eq:cov_matrix_eig}
\end{equation}
and $\theta_{j,j'}=[ {\cal C} ( t \rightarrow \infty) ]_{j,j'}$.
\noindent
In panel c) of Fig.\ref{fig:BA2_typeIII_single}, we show the distribution of $\nu_{\epsilon}$ for a single disorder configuration of $H_2^{B-A}$ with $\sigma=0.5$ and $W=1.8$. We have set the hopping strengths values to $v=1.25$ and $w=1.7$, corresponding to an EOD of $\bar{\nu}=-2.351$. 
Although peaked around zero, $\nu_{\epsilon}$ takes negative and positive values, with zero mean. In real space, the states associated with  $\nu_{\epsilon}<0$ are localized on the left hand side of the eSSH chain while the states with $\nu_{\epsilon}>0$ are on its right hand side. The first has an occupation probability $\theta_{\epsilon}\approx1$, being connected to the bath that injects electrons, while the others have $\theta_{\epsilon}\approx0$, being localized near the sink bath. It follows that the total EOD is the sum of the contribution given by all the states localized near the left edge. Each of these states gives a small contribution to the overall EOD up to the observed value $\bar{\nu}=-2.351$.
This behavior is totally different from that observed in the topological phase where the number of eigenstates with $\nu_{\epsilon}\neq0$ is expected to be equal to twice the value of the EOD, half of them localized on the left edge and the other half on the right one, as shown in panel d) of Fig.\ref{fig:BA2_typeIII_single}. For $v=0.25$, $w=0.5$, $\sigma=0.5$ and $W=0.6$, for Type I disorder the system is in the topological phase with EOD equal to $2$. Indeed, only four states with non zero $\nu_{\epsilon}$ appears in the histogram, where the states with positive (negative) EOD are localized on the left (right) side.

In the topological phase, all the eigenstates except for the zero energy edge states occupy the A and B sublattices with the same probability weight. The edge states show instead a preference to lay on only one of the sublattices as a function of the sign of the topological invariant. The number of states laying on the A (B) sublattice on the left edge is equal to the EOD if it is positive (negative) and vice versa on the right hand side.
In the NCCG phase, there is still a trend on the part of the eigenstates to prefer the A or B sublattice on each side of the chain as a function of the sign of the EOD. However,rather than involving a number of states exactly equal to twice the value of the EOD, the total contribution is split between many nearly localized non topological states each of them carrying a small fraction of the overall EOD.

\section{Conclusions}
\label{sec:conclusions}

We have studied different eSSH models, i.e., SSH models with long range hopping amplitudes, by means of the LE formalism. We have shown that the effect of correlated and non-correlated disorder preserving the chiral symmetry or not is very different.
By inducing the system into a NESS, coupling the system to two external baths in the large bias regime, we have discussed how the EOD and its standard deviation can be used to track the fate of the topological phases as a function of the disorder strength. In the presence of disorder that preserves the chiral symmetry, the topological phases characterized by an higher integer value of the EOD are hierarchically destroyed in favor of phases with a lower value of the EOD. In the process, disorder induced "buffer" phases, separated from each other by a Griffits region and characterized by zero standard deviation, are introduced so that the EOD decreases at unitary steps. Phases with EOD=$\pm 1$ are more robust to disorder and can be enhanced by strong disorder. On the contrary, the topological phase is lost if disorder breaks the chiral symmetry and a new phase, characterized by a non integer EOD and a large standard deviation, emerges.
While, to illustrate the application of our method, in this paper we limited ourselves to some particular kinds of disorder and eSSH models, there are no limitations to apply our method to eSSH models with more than one long range term or mixture of different types of disorder.
We expect that our findings can be observed experimentally in ultracold atoms \cite{Schaff2010,Billy2008,Sanchez-Palencia2010} and photonic systems \cite{Dietz2011}, where the EOD can be implemented as a tool to investigate the robustness of a given topological phase, and the corresponding zero energy modes, as a function of the system parameters and disorder strength. Other possible applications should concern, for instance, novel topological phases/phase transitions arising in the phase diagram of junctions of interacting fermionic systems and/or spin chains \cite{Tsvelik2013,Guerci2021,Giuliano2020,Giuliano2020_2,Giuliano2022,Buccheri2022}.

\vspace{0.3cm} 
 
{\bf Acknowledgements:}   We thank R. Egger for insightful discussions.  
 A.N. acknowledge  funding by the Deutsche Forschungsgemeinschaft (DFG, German Research Foundation) under Grant No.~ 277101999, TRR 183 (project C01), under Germany's Excellence Strategy - Cluster of Excellence Matter and Light for Quantum Computing (ML4Q) EXC 2004/1 - 390534769, and under Grant No.~EG 96/13-1. E.G.C. acknowledge funding (partially) supported by ICSC – Centro Nazionale di Ricerca in High Performance Computing, Big Data and Quantum Computing, funded by European Union – NextGenerationEU. We acknowledge computer resources from "PIR01-00011 “IBISCo”, PON 2014-2020".

\appendix 

\section*{Appendix: Derivation of the recursive SDRG equation}
\label{app:SDRG}
In this Appendix we perform the explicit calculations reviewed in Section \ref{sec:SDRG} to derive both the closed formula presented in Eq.\eqref{eq:transition_line} for the SSH model and the more general recursive formula of Eq.\eqref{eq:RG_relation} for a generic long range SSH model (including the eSSH models discussed in this paper).

\subsection{SDRG applied to the disordered SSH model}
\label{app:SDRG_SSH}
Starting from the SSH Hamiltonian defined in Eq.\eqref{eq:ssh_normal}, we regard $v$ and $w$ as site-dependent random numbers. We therefore rewrite the main Hamiltonian as
\begin{equation}
    H_{v,w}=\sum_{j=1}^{N}\left( v_jc^\dagger_{A,j}c_{B,j}+w_j c^\dagger_{B,j} c_{A,j+1}\right)+\text{h.c.}
    \label{eq:SSH_normal_perturbed}
\end{equation}
where $v_i$ and $w_i$ are real, positive parameters and come from two probability distributions that can, in principle, be different.
We now set ${\Omega=\max(\{v_i\},\{w_i\})=v_k}$ and thus isolate the contribution in Eq.\eqref{eq:SSH_normal_perturbed} that depends on this hopping
\begin{equation}
    H_{k}=v_k\left( c^\dagger_{A,k}c_{B,k}+c^\dagger_{B,k} c_{A,k}\right)
\end{equation}
It is thus possible to project $H_{v_k}$ in the subspace generated by the basis vector associated with the sites $(A,k)$ and $(B,k)$, ordered as
\begin{equation}
    \{\ket{i_{A,k},i_{B,k}}\}=\{\ket{0,0},\ket{0,1},\ket{1,0},\ket{1,1}\}    
\end{equation}
where
\begin{equation}
    \ket{1,1}=c^\dagger_{A,i}c^\dagger_{B,i}\ket{0,0}
\end{equation}
In this basis, the Hamiltonian is a $4 \times 4$ matrix
\begin{equation}
\bra{i_{A,k},j_{B,k}}H_{k}\ket{m_{A,k},n_{B,k}}=v_k
	\begin{pmatrix} 
		0 & 0 & 0 & 0 \\ 
		0 & 0 & 1 & 0 \\ 
		0 & 1 & 0 & 0 \\ 
		0 & 0 & 0 & 0
	\end{pmatrix}
\end{equation}
The eigenvalues and eigenstates are
\begin{center}
    \begin{tabular}{l|l}
    Eigenvalue & Eigenstate \\
    \hline
         $E_{1,\pm}=\pm v_k$& $\ket{\psi_{1,\pm}}=\frac{1}{\sqrt{2}}(c^\dagger_{B,k}\pm c^\dagger_{A,k})\ket{0,0}$\\
         \hline
         \multirow{ 2}{*}{$E_{0,\pm}=0$}            & $\ket{\psi_{0,-}}=\ket{0,0}$ \\ & $\ket{\psi_{0,+}}=c^\dagger_{A,k}c^\dagger_{B,k}\ket{0,0}$
    \end{tabular}
\end{center}
Since $v_k\ge 0$, the local ground state is $\ket{\psi_{1,-}}$ with energy $E_{1,-}=-v_k$. Projecting the Hamiltonian in Eq.\eqref{eq:SSH_normal_perturbed} into the subspace generated by this state, we obtain an effective Hamiltonian having two less degrees of freedom (i.e., deprived of the $k$-th dimer)
\begin{equation}
H_{v,w}^{\mathrm{eff}}=H_{v,w}-H_k-V+\Delta
\label{eq:effective_H}
\end{equation}
where $V$ is the part of the Hamiltonian coupling the $k$-th dimer with the rest of the chain, namely,
\begin{align}
    V=& w_{k-1}\left( c^\dagger_{B,k-1} c_{A,k} + c^\dagger_{A,k}c_{B,k-1}\right) \\       & +w_{k}\left( c^\dagger_{B,k} c_{A,k+1} + c^\dagger_{A,k+1}c_{B,k}\right)           \nonumber
\end{align}
and $\Delta$ is the sum of the local ground-state energy and the perturbation expansion of $V$ with respect to the local ground state subspace. Up to second-order, we have
\begin{align}
\Delta & =  E_{1,-} \nonumber\\
         & + \bra{\psi_{1,-}}V\ket{\psi_{1,-}} \nonumber \\
         & + \sum_{n,\nu}                 \frac{\bra{\psi_{1,-}}V\ket{\psi_{n,\nu}}\bra{\psi_{n,\nu}}V\ket{\psi_{1,-}}}{E_{1,-}-E_{n,\nu}}
\end{align}
%The next step is to calculate $\Delta$ explicitly: in order to achieve this result, first of all one calculates the matrix elements of $V$ using the local eigenstates. It is worth noticing that the matrix elements will be operators and not simple numbers, because the projection is only done on the local Hilbert space:
with the matrix element of $V$ given by
\begin{align}
    & \bra{\psi_{n,\pm}}V\ket{\psi_{n,\nu}}=0 \quad n=0,1\quad\nu=\pm \\
    & \bra{\psi_{0,-}}V\ket{\psi_{1,\nu}}=\frac{1}{\sqrt{2}}\Big[w_k c^\dagger_{A,k+1}+\nu w_{k-1}c^\dagger_{B,k-1}\Big] \\
    & \bra{\psi_{0,+}}V\ket{\psi_{1,\nu}}=\frac{1}{\sqrt{2}}\Big[\nu w_k c_{A,k+1}- w_{k-1}c_{B,k-1}\Big]
\end{align}
%Since the diagonal contributions are zero, the first nonzero perturbative contribution is of the second order. Its explicit form is
It follows that
\begin{align}
    \Delta = & -\frac{w_{k-1}w_k}{v_k}\left(c^\dagger_{B,k-1}c_{A,k+1}+c^\dagger_{A,k+1}c_{B,k-1}\right) \nonumber\\
    & -v_k -\frac{w_{k-1}^2+w_{k}^2}{2v_k}
    \label{eq:SDRG_perturb_SSH}
\end{align}
%\begin{align}
%    \Delta
%    = & -v_k + \sum_{n,\nu} %\frac{\bra{\psi_{1,-}}V\ket{\psi_{n,\nu}}\bra{\psi_{n,\nu}}V\ket{\psi_{1,-}%}}{E_{1,-}-E_{n,\nu}} \\
%    = & -v_k + \sum_{\nu} %\frac{\bra{\psi_{1,-}}V\ket{\psi_{0,\nu}}\bra{\psi_{0,\nu}}V\ket{\psi_{1,-}%}}{E_{1,-}-E_{0,\nu}} \nonumber\\
%    = & -v_k + \frac{1}{E_{1,-}}\sum_{\nu}\bra{\psi_{1,-}}V\ket{\psi_{0,\nu}}\bra{\psi_{0,\nu}}V\ke%t{\psi_{1,-}} \nonumber\\
%    = & -\frac{w_{k-1}w_k}{v_k}\left(c^\dagger_{B,k-1}c_{A,k+1}+c^\dagger_{A,k+1}c_{B,k-1}\right) \nonumber\\
%    & -v_k -\frac{w_{k-1}^2+w_{k}^2}{2v_k} \nonumber
%    \label{eq:SDRG_perturb_SSH}
%\end{align}
%Plugging back this to \eqref{eq:effective_H}, we obtain:
%\begin{align}
%H_{v,w}^{eff}  = & H_{v,w}-H_k-V+\Delta \nonumber \\
% =& \sum_{j\ne k} v_j\left( c^\dagger_{A,j}c_{B,j}+c^\dagger_{B,j} %c_{A,j}\right) \nonumber \\
% +& \sum_{j\ne k,k-1} w_{j}\left( c^\dagger_{B,j} c_{A,j+1} + %c^\dagger_{A,j+1}c_{B,j}\right) \nonumber \\
% - &\frac{w_{k-1}w_k}{v_k}\left(c^\dagger_{B,k-%1}c_{A,k+1}+c^\dagger_{A,k+1}c_{B,k-1}\right) \nonumber \\
% - &v_k -\frac{w_{k-1}^2+w_{k}^2}{2v_k}
%\end{align}
Apart from a constant shift, the second order contribution renormalizes the original Hamiltonian $H_{v,w}$ into a new effective one without the $(A,k)$ and $(B,k)$ sites, and with sites $(B,k-1)$ and $(A,k+1)$ connected by a new effective coupling $\tilde{w_k}=-\frac{w_{k-1}w_k}{v_k}$.
\\
Similarly, if ${\Omega=\max(\{v_i\},\{w_i\})=w_k}$, we remove the $(B,j)$ and $(A,k+1)$ sites so that the $(A,j)$ and $(B,k+1)$ ones are connected by a new effective coupling $\tilde{v_k}=-\frac{v_{k-1}v_k}{w_k}$.
\\
After performing $l$ renormalization steps on each different coupling, their values are given by
\begin{align}
    & \tilde{v}= \frac{v_kv_{k+1}\dots v_{k+l}}{w_kw_{k+1}\dots w_{k+l}} \\
    & \tilde{w}= \frac{w_kw_{k+1}\dots w_{k+l}}{v_kv_{k+1}\dots v_{k+l}}
\end{align}
If $l\gg 1$, we can perform the change of variable $x=e^{\ln{x}}$ and retrieve equation \eqref{eq:DDRG_asymp_coupling}
\begin{align}
    & |\tilde{v}|\xrightarrow[]{l\gg 1} e^{l(\left<\ln v\right> - \left<\ln w\right>)} \\
    & |\tilde{w}|\xrightarrow[]{l\gg 1} e^{l(\left<\ln w\right> - \left<\ln v\right>)}
\end{align}
The transition line given by this scheme is the one by which neither $\tilde{v}$ nor $\tilde{w}$ flows toward a zero value, i.e., when the nondiverging part of the exponent is zero
\begin{equation}
    \left<\ln v\right> = \left<\ln w\right>
\end{equation}
retrieving Eq.\eqref{eq:transition_line}.

\subsection{SDRG for the eSSH in the presence of disorder preserving the chiral symmetry}
Let us apply the SDRG scheme in the presence of long range hoppings that preserve the chiral symmetry, i.e., Eq.\eqref{eq:generic_H}
\begin{equation}
    \mathcal{H}=\sum_{ij}K_{ij}\left(c^\dagger_{A,i}c_{B,j} + c^\dagger_{B,j}c_{A,i}\right)
    \label{eq:long_range_H_chiral}
\end{equation}
where the couplings $K_{ij}$ are generated from one or more random distributions. Let us assume, without loss of generality, that $\Omega=\max({\{|K_{ij}|\}})=K_{lm}$ with $l<m$. We isolate the part of $\mathcal{H}$ depending on this parameter
\begin{equation}
    \mathcal{H}_{lm}=K_{lm}\left(c^\dagger_{A,l}c_{B,m} + c^\dagger_{B,m}c_{A,l}\right)
\end{equation}
and compute its eigenvalues and eigenvectors
\begin{center}
    \begin{tabular}{l|l}
    Eigenvalue & Eigenstate \\
    \hline
         $E_{1,\pm}=\pm K_{lm}$& $\ket{\psi_{1,\pm}}=\frac{1}{\sqrt{2}}(c^\dagger_{A,l}\pm c^\dagger_{B,m})\ket{0_{A,l},0_{B,m}}$\\
         \hline
         \multirow{ 2}{*}{$E_{0,\pm}=0$}            & $\ket{\psi_{0,-}}=\ket{0_{A,l},0_{B,m}}$ \\
          & $\ket{\psi_{0,+}}=c^\dagger_{A,l}c^\dagger_{B,m}\ket{0_{A,l},0_{B,m}}$\end{tabular}
\end{center}
As it was done in the previous Section, we define an effective Hamiltonian by projecting the full Hamiltonian in Eq.\eqref{eq:long_range_H_chiral} on the local ground state $\ket{\psi_{1,-}}$, and by treating the terms of the Hamiltonian that depend on $c^\dagger_{A,l}$, $c_{A,l}$, $c^\dagger_{B,m}$, and $c_{B,m}$ as a perturbation. More explicitly, this means that 
\begin{equation}
    \mathcal{H}^{\mathrm{eff}}=\mathcal{H}-\mathcal{H}_{lm}-\mathcal{V}+\Delta
\end{equation}
with $\mathcal{V}$ given by
\begin{align}
    \mathcal{V}= & \sum_{i\ne m}K_{li}\left(c^\dagger_{A,l}c_{B,i}+ c^\dagger_{B,i}c_{A,l}\right) \nonumber \\ +& \sum_{i\ne l}K_{im}\left(c^\dagger_{A,i}c_{B,m} + c^\dagger_{B,m}c_{A,i}\right)
\end{align}
and $\Delta$, up to second order, given by
\begin{align}
\Delta & =  E_{1,-} \nonumber\\
         & + \bra{\psi_{1,-}}\mathcal{V}\ket{\psi_{1,-}} \nonumber \\
         & + \sum_{n,\nu}                 \frac{\bra{\psi_{1,-}}\mathcal{V}\ket{\psi_{n,\nu}}\bra{\psi_{n,\nu}}\mathcal{V}\ket{\psi_{1,-}}}{E_{1,-}-E_{n,\nu}}
\end{align}
with the matrix elements of $\mathcal{V}$ equal to
\begin{align}
    & \bra{\psi_{n,\pm}}\mathcal{V}\ket{\psi_{n,\nu}}=0 \quad n=0,1 \quad \nu=\pm \\
    & \bra{\psi_{0,-}}\mathcal{V}\ket{\psi_{1,\nu}}=\frac{1}{\sqrt{2}}\sum_i\left[\left(1-\delta_{im}\right) K_{li}c^\dagger_{B,i}+\right. \\
    & \qquad\qquad\qquad\qquad\qquad\left. +\nu \left(1-\delta_{il}\right) K_{im}c^\dagger_{A,i}\right] \nonumber \\
    & \bra{\psi_{0,+}}\mathcal{V}\ket{\psi_{1,\nu}}=\frac{1}{\sqrt{2}}\sum_i\left[\left(1-\delta_{il}\right) K_{im}c^\dagger_{A,i}+\right. \\ 
    & \qquad\qquad\qquad\qquad\qquad\left.-\nu \left(1-\delta_{im}\right) K_{li}c^\dagger_{B,i}\right] \nonumber
\end{align}
%As for the SSH case, the matrix elements are operators too and as such they obey fermionic anti-commutation rules.
%Since the first order perturbative term is zero because the diagonal terms are zero, we calculate the second order perturbative contribution
It follows that
\begin{align}
%    \Delta
%     = &-K_{l,m}+\sum_{n,\nu} \frac{\bra{\psi_{1,-}}\mathcal{V}\ket{\psi_{n,\nu}}\bra{\psi_{n,\nu}}\mathcal{V}\ket{\psi_{1,-}}}{E_{1,-}-E_{n,\nu}} \\
%     = &-K_{l,m}-\frac{1}{E_{1,-}}\sum_{\nu}\bra{\psi_{1,-}}\mathcal{V}\ket{\psi_{0,\nu}}\bra{\psi_{0,\nu}}\mathcal{V}\ket{\psi_{1,-}} \nonumber\\
    \Delta = & -\sum_{ij}\left(1-\delta_{il}\right)\left(1-\delta_{jm}\right)\frac{K_{lj}K_{im}}{K_{lm}} \left(c^\dagger_{A,i}c_{B,j}+c^\dagger_{B,j}c_{A,i}\right) \nonumber\\
       & -\sum_{i}\left(1-\delta_{il}\right)\left(1-\delta_{jm}\right)\frac{K_{li}^2+K_{im}^2}{2K_{lm}}-K_{l,m} 
\end{align}
Apart from an overall shift, the effective Hamiltonian has two less sites, and the couplings are renormalized through the following relation
\begin{equation}
    \tilde{K}_{ij}=K_{ij}-\frac{K_{lj}K_{im}}{K_{lm}}\quad i\ne l,\text{  } j\ne m
    \label{eq:recursive_long}
\end{equation}
Unlike the simple SSH chain, in the long range SSH model it is not easy to analytically iterate Eq.\eqref{eq:recursive_long} to retrieve a closed formula. The asymptotic flow of the hopping terms should be recovered numerically, iterating Eq.\eqref{eq:recursive_long}, as discussed in the main text.
\bibliography{biblio_essh}
\end{document}